\documentclass[twocolumn,         
  showpacs,longbibliography,            
  showkeys,preprintnumbers,     
  aps,                 
  prd,                
  letterpaper,             
  nofootinbib,         
  tightenlines,        
  floats,floatfix      
]{revtex4-1}
\usepackage{physics}
\usepackage{epsf}
\usepackage{graphicx,color}
\usepackage{subfigure}
\usepackage{latexsym}
\usepackage{amsmath,amssymb}
\usepackage[colorlinks=true,linkcolor=blue,citecolor=blue]{hyperref}
\usepackage{mathrsfs}
\usepackage{comment}
\usepackage{soul}
\usepackage{cancel}
\definecolor{purple}{rgb}{0.58,0.0,0.83}
\definecolor{orange}{rgb}{1,0.5,0}
\DeclareSymbolFontAlphabet{\mathrsfs}{rsfs}
\DeclareMathAlphabet{\mathcal}{OMS}{cmsy}{m}{n}

\newcommand\fb[1]{{\color{blue}  #1}}

\begin{document}

\thispagestyle{empty}

\title{Accretion of multipolar massive complex scalar field packets by a Schwarzschild black hole}

\author{Flavio Rosales-Infante}
\email{flavio.rosales.infante@umich.mx}
\affiliation{Instituto de F\'{\i}sica y Matem\'{a}ticas, Universidad
  Michoacana de San Nicol\'as de Hidalgo. Edificio C-3, Cd.
  Universitaria, 58040 Morelia, Michoac\'{a}n,
M\'{e}xico.}

\author{Ivan Álvarez-Ríos}
\email{ivan.alvarez@umich.mx}
\affiliation{Instituto de F\'{\i}sica y Matem\'{a}ticas, Universidad
  Michoacana de San Nicol\'as de Hidalgo. Edificio C-3, Cd.
  Universitaria, 58040 Morelia, Michoac\'{a}n,
M\'{e}xico.}

\author{Francisco S. Guzm\'an}
\email{francisco.s.guzman@umich.mx}
\affiliation{Instituto de F\'{\i}sica y Matem\'{a}ticas, Universidad
  Michoacana de San Nicol\'as de Hidalgo. Edificio C-3, Cd.
  Universitaria, 58040 Morelia, Michoac\'{a}n,
M\'{e}xico.}

\begin{abstract}
We study the finite-time accretion of complex massive scalar wave packets by a Schwarzschild black hole in the test-field regime, with parameters motivated by ultralight fuzzy dark matter around supermassive black holes. Our goal is to determine how the scalar content of a localized configuration is redistributed after interacting with the black hole, and which spectral and multipolar components are more efficiently absorbed. We decompose the Klein--Gordon field into independent multipolar sectors and evolve nearly monochromatic Gaussian packets mode by mode, reducing the problem to a set of 1+1 dimensional evolutions. Accretion is quantified with the flux of the conserved Noether current through the horizon surface, providing a direct measure of the scalar charge absorbed by the black hole. For a carrier radial wavenumber $k_0$ and multipole index $\ell$, we construct accretion-efficiency maps in the $(k_0,\ell)$ plane that contain the fraction of accreted modal charge. These maps exhibit a transition between inefficient, partial, and efficient accretion regimes, which we relate to the structure of an effective potential. We show that the process is controlled by the ratio between the Schwarzschild radius $R_s$ and the reduced Compton wavelength $\lambdabar_C$. For $R_s \lesssim \lambdabar_C$, the transition is broad and dominated by the angular momentum barrier, while for $R_s > \lambdabar_C$ it sharpens across a narrower range of $k_0$ and a partial-accretion floor emerges at low $k_0$. These results provide a time-domain, Noether-charge-based classification of black hole accretion for massive scalar wave packets.
\end{abstract}

\maketitle

\section{Introduction}
\label{sec:introduction}

Massive scalar fields around black holes are an interesting scenario to study how strong gravity redistributes wave like matter. In a Schwarzschild spacetime, a scalar field configuration does not interact with the horizon as a set of point-like particles, instead its radial scale, angular structure, and field mass determine how much of the scalar matter content crosses the horizon and how much remains outside. This separation is determined by the geometry of the space-time,  wave velocities, angular distribution and mass  of the field. 

The standard description of this scenario is based on the analysis of stationary scattering. In this framework, the black hole response is characterized by reflection and transmission coefficients, greybody factors, and absorption cross sections. The foundational analyses of scalar absorption by Schwarzschild black holes established that the response depends on the relation between the incident wavelength and the black hole radius as found by the essential results by Unruh and Fabbri \cite{Sanchez1976,Fabbri1975,Unruh1976}. Later developments refined this barrier transmission picture through low frequency universality results and phase integral descriptions \cite{DasGibbonsMathur1997,Andersson1994}. Similar scattering and absorption analyses have also been extended to charged, rotating, and other space-time geometries \cite{NakamuraSato1976,BenoneOliveiraDolanCrispino2014, LeiteBenoneCrispino2017,LimaJuniorBenoneCrispino2020}.

Stationary quantities provide the asymptotic response of the system, whereas spatially localized scalar configurations raise a complementary question. Given a finite packet of scalar field initially placed outside the black hole, one can ask how its density is distributed after the interaction, specifically which part crosses the horizon, and which part remains as an exterior remnant. This question can be answered by following the evolution of a scalar field packet that interacts with the black hole.


Because of its potential astrophysical relevance, in our analysis we consider a complex massive scalar field. The global $U(1)$ symmetry of this field provides a conserved Noether current, and the associated charge gives a direct estimate of the scalar content carried by a wave packet. In an accretion scenario, like that involving a black hole, the flux through the horizon of this current serves to measure the accreted charge, while the charge remaining outside the black hole measures the exterior remnant. 

Now, the scalar field mass is important because it introduces an intrinsic length scale, the reduced Compton wavelength $\lambdabar_C$. Previous studies based on numerical simulations with spherical symmetry, indicate that this scale is fundamental to determine whether a scalar field is totally or partially accreted, for example in a fixed background \cite{scri} or in full non-linear numerical simulations \cite{GuzmanLora2012,PhysRevD.104.084014}, the Compton wavelength is essential. In more general scalar field configurations the black hole response would then be controlled not only by the radial wavenumber scale and the angular structure of the field, but also by the relation between the Schwarzschild radius $R_s$ and $\lambdabar_C$. 

Perhaps one of the most astrophyscally relevant scenarios is that of ultralight bosonic dark matter \cite{Matos:2000ss,Hui:2016,Niemeyer_2020}, which shows wave-like behavior at galactic scales due to the prediction of galactic cores, but is precisely the accretion onto a Supermassive Black Hole that becomes a scenario that could test this dark matter candidate, specially when the accretion rate or the Compton length scale could determine accretion rates. When the dark matter particle is ultralight with masses of order $10^{-23}-10^{-19}{\rm eV}/{\rm c^2}$, accretion or remnants are crucial, which is why in this paper our numerical examples use scalar field mass within this range. 

In order to study the accretion, scattering and retention of the scalar field, in this paper we evolve multi-mode, nearly monochromatic Gaussian wave packets with radial wavenumber $k_0$ on a fixed Schwarzschild background and measure their Noether charge mode by mode. We use the fact that the spherical symmetry of the spacetime allows the field to be decomposed into independent multipolar sectors, each one defining an evolution problem characterized by the pair $(k_0,\ell)$. For each sector, we calculate the accreted fraction of the scalar charge and use it to construct black hole response maps in the $(k_0,\ell)$ plane. These maps allow us to classify the fate of the scalar field into efficiently accreted, partially accreted, and exterior retained regions. We then compare the structure observed in the maps with the classical transmission threshold determined by the maximum of the massive scalar effective potential, and analyze the response in the regimes $R_s \leq \lambdabar_C$ and $R_s > \lambdabar_C$. In this sense, the present work provides a charge based, time domain classification of effective accretion for massive scalar wave packets and addresses which multipolar components of a localized massive scalar configuration are absorbed by a Schwarzschild black hole, which remain outside, and how this separation is controlled by $k_0$, $\ell$, and the scalar field mass.

The paper is organized as follows. In Sec.~\ref{sec:covariant_formulation} we introduce the covariant Klein-Gordon dynamics, the global $U(1)$ current, and the Noether charge accretion diagnostic. In Sec.~\ref{sec:modal_reduction} we present the first order formulation and the multipolar reduction on the Schwarzschild background. In Sec.~\ref{sec:numerical_approach} we describe the numerical method and the validation tests. In Sec.~\ref{sec:mono_waves} we present representative evolutions and accretion efficiency maps for nearly monochromatic Gaussian scalar wave packets. Finally, in Sec.~\ref{sec:conclusions} we summarize the main results.

\section{Covariant formulation and conserved dynamics}
\label{sec:covariant_formulation}

\subsection{3+1 description of the complex scalar field dynamics}

We consider a complex scalar field $\phi$ propagating on a fixed curved spacetime background, whose dynamics is governed by the action

\begin{equation}
  S[\phi,\phi^*]  =  \int d^4x\,\sqrt{-g}  \left(    - g^{\mu\nu}\nabla_\mu\phi^*\,\nabla_\nu\phi    - V(\phi)\right), 
  \label{eq:action}
\end{equation}

\noindent where $g^{\mu\nu}$ is the contravariant metric tensor, $\nabla_\mu$ is the covariant derivative compatible with the spacetime metric, and $V(\phi)$ is the scalar potential \cite{Wald1984,BirrellDavies1982}. In this work we consider a free field without self-interaction, which corresponds to the potential

\begin{equation}
  V(\phi)=\mu^2\phi^*\phi,\nonumber
\end{equation}

\noindent where $\mu$ is the mass parameter of the scalar field, defined as $\mu=1/\lambdabar_C$, with $\lambdabar_C=\hbar/(m_b c)$ the reduced Compton wavelength associated with a scalar field particle of rest mass $m_b$.

The variation of the action (\ref{eq:action}) with respect to $\phi^*$ yields the Klein-Gordon (KG) equation

\begin{equation}
  \Box\phi-\mu^2\phi=0,
  \label{eq:KG_covariant}
\end{equation}

\noindent where $\Box=\nabla^\mu\nabla_\mu$ is the covariant d'Alembert operator associated with the spacetime metric. For the description of the spacetime geometry, we use the $3+1$ decomposition with the line element \cite{Alcubierre2008}:

\begin{equation}
  ds^2  =  -\alpha^2 c^2 dt^2 +  \gamma_{ij}\left(dx^i+\beta^i c\,dt\right)\left(dx^j+\beta^j c\,dt\right), \nonumber
\end{equation}

\noindent where $\alpha$ is the lapse function, $\beta^i$ the shift vector, and $\gamma_{ij}$  the induced spatial 3-metric on hypersurfaces $\Sigma_t$ of constant coordinate time $t$. The future-directed unit normal vector to these hypersurfaces $n^\mu$ is defined by the conditions 

\begin{equation}
  n_\mu n^\mu=-1,  \qquad   n_\mu dx^\mu=-\alpha c\,dt,\nonumber
\end{equation}

\noindent which explicitly is written as

\begin{equation}
  n_\mu=(-\alpha c,0,0,0),  \qquad  n^\mu=\left(\frac{1}{\alpha c},-\frac{\beta^i}{\alpha}\right),\nonumber
\end{equation}

\noindent in coordinates adapted to the 3+1 foliation.

\subsection{Global $U(1)$ symmetry and Noether current}

The complex KG field admits a global phase symmetry,

\begin{equation}
  \phi \mapsto e^{i\chi}\phi,  \qquad  \chi=\mathrm{const},\nonumber
\end{equation}

\noindent and by Noether's theorem, there is an associated conserved current. For the above invariance, the corresponding Noether current is

\begin{equation}
  J^\mu \equiv \mathrm{Im}\!\left(\phi^* \nabla^\mu \phi\right),\nonumber
\end{equation}

\noindent which is associated with rotations in the complex plane of the field \cite{BirrellDavies1982,PeskinSchroeder1995}. For solutions of Eq.~\eqref{eq:KG_covariant}, this current satisfies the conservation law

\begin{equation}
  \nabla_\mu J^\mu=0,
  \label{eq:conserv_covariant}
\end{equation}

\noindent where the current is related to the particle-number density measured by Eulerian observers 

\begin{equation}
  \rho_n\equiv -n_\mu J^\mu,\nonumber
\end{equation}

\noindent whose integral over a spatial hypersurface $\Sigma_t$ gives the total particle number at a given coordinate time $t$

\begin{equation}
  N(t)=\int_{\Sigma_t}\sqrt{\gamma}\,\rho_n\,d^3x,
  \label{eq:Noether_charge}
\end{equation}

\noindent where $\gamma$ is the determinant of $\gamma_{ij}$. In the $3+1$ decomposition above, Eq.~\eqref{eq:conserv_covariant} takes the following flux-balance form

\begin{equation}
  \partial_t\!\left(\sqrt{\gamma}\,\rho_n\right)  +  \partial_i\!\left(\alpha c\, \sqrt{\gamma}\,J^i\right)  =  0,
  \label{eq:continuity}
\end{equation}

\noindent which states that local variations of the particle number contained in a spatial region are determined by the flux of the spatial current through its closed boundary surface. This balance law provides the basis for the accretion, since it gives a direct measure of how much conserved scalar charge crosses the black-hole horizon surface.

\subsection{Relativistic Madelung decomposition}

An amplitude-phase representation of wave dynamics was originally introduced in the non-relativistic context by Madelung \cite{Madelung1926,Bohm1952}. For relativistic scalar fields, the same construction provides an exact rewriting of the KG equation in terms of amplitude and phase variables \cite{Wong2010}, where the complex field is written as

\begin{equation}
  \phi=\sqrt{\varrho}\,e^{iS/\hbar},\nonumber
\end{equation}

\noindent $\varrho\ge 0$ is the squared amplitude of the scalar field, and $S$ is a real phase. Substituting this expression into Eq.~\eqref{eq:KG_covariant} and separating real and imaginary parts gives two coupled equations \cite{Wong2010}:

\begin{eqnarray}
  &&\nabla_\mu\!\left(\frac{\varrho}{\hbar}\nabla^\mu S\right)=0,
  \label{eq:Madelung_continuity_covariant}\\
  &&\nabla_\mu S \nabla^\mu S  + \hbar^2 \mu^2  -  \hbar^2 \frac{\Box\sqrt{\varrho}}{\sqrt{\varrho}}  =  0.
  \label{eq:Madelung_HJ_covariant}
\end{eqnarray}

\noindent The first of these equations is the conservation law associated with the global $U(1)$ symmetry \eqref{eq:conserv_covariant}, since in Madelung variables the Noether current becomes

\begin{equation}
  J^\mu=\frac{\varrho}{\hbar}\,\nabla^\mu S.\nonumber
\end{equation}

\noindent The continuity equation obtained from the Madelung decomposition and the conservation law obtained from Noether's theorem are equivalent expressions of the same conserved current. This equivalence is useful because it is the Noether theorem in terms of the amplitude and phase of the field. The Second equation, Eq. \eqref{eq:Madelung_HJ_covariant} determines the phase evolution and has the form of the relativistic Hamilton-Jacobi equation for the phase, corrected by the term

\begin{equation}
  Q\equiv \hbar^2 \frac{\Box\sqrt{\varrho}}{\sqrt{\varrho}},\nonumber
\end{equation}

\noindent which may be interpreted as a relativistic quantum potential. When this term is negligible, the phase gradient satisfies the standard massive Hamilton-Jacobi relation for the phase $\nabla_\mu S \nabla^\mu S + \hbar^2 \mu^2 = 0$.

In order to characterize the phase we introduce the phase covector

\begin{equation}
  k_\mu \equiv \frac{1}{\hbar} \nabla_\mu S,\nonumber
\end{equation}

\noindent whose temporal projection defines the local frequency, while its spatial projection defines the spatial wave covector as follows:

\begin{equation}
  \omega \equiv -\frac{c}{\hbar}\,n^\mu\nabla_\mu S,  \qquad  k_i\equiv \frac{1}{\hbar} D_i S,   \label{eq:phase_frequency_wavenumber}
\end{equation}

\noindent where $D_i$ is the covariant derivative on the spatial hypersurface $\Sigma_t$ with metric $\gamma_{ij}$. Thus, $\omega$ is the angular frequency measured by Eulerian observers, while $k_i$ is the spatial wave covector on $\Sigma_t$. The phase covector can be decomposed as

\begin{equation}
  \frac{1}{\hbar} \nabla_\mu S  =  \frac{\omega}{c}\,n_\mu + k_\mu,\nonumber
\end{equation}

\noindent with the condition $ k_\mu n^\mu = 0$, which means that the wave covector is purely spatial. Substituting this split into Eq.~\eqref{eq:Madelung_HJ_covariant} gives

\begin{equation}
  -\frac{\omega^2}{c^2}  +  k^i k_i + \hbar^2 \mu^2  -\frac{Q}{\hbar^2}  =  0,\nonumber
\end{equation}

\noindent whereas the Eulerian density and the spatial components of the conserved current become

\begin{equation}
  \rho_n = \frac{\varrho\,\omega}{c},
~~~~  J^i  =  \varrho \left(
    \gamma^{ij}k_j    -    \frac{\beta^i}{\alpha c}\,\omega
  \right).\nonumber
\end{equation}

\subsection{Schwarzschild background and accretion rate}

We describe the Schwarzschild space-time with ingoing Eddington--Finkelstein coordinates, which are regular across the event horizon and are suitable for the study of accretion \cite{BarrancoEtAl2011,GuzmanLora2012,LoraClavijoGraciaLinaresGuzman2014}. In these coordinates, the lapse function, shift vector, and spatial metric take the form

\begin{eqnarray}
    \alpha &=& \left(1+\frac{R_s}{r}\right)^{-1/2}, \nonumber\\
    \beta^i &=& \left(\frac{R_s/r}{1+R_s/r},0,0\right), \nonumber\\
    \gamma_{ij} &=& \mathrm{diag}\left(1+\frac{R_s}{r},r^2,r^2\sin^2\theta\right),\nonumber
\end{eqnarray}

\noindent where $R_s=2GM/c^2$ is the Schwarzschild radius of a black hole of mass $M$. For this background, the radial component of the current reads

\begin{equation}
  J^r  =  \varrho  \left(
    \gamma^{rr}k_r    -    \frac{\beta^r}{\alpha c}\,\omega  \right),\nonumber
\end{equation}

\noindent which depends on both the spatial phase gradient $k_r$ and the temporal projection of the phase covector $\omega$ defined in Eq. \eqref{eq:phase_frequency_wavenumber}.

Now, the particle-number accretion rate by the black hole is defined as the radial flux evaluated at the horizon surface with radius  $r=R_s$:

\begin{equation}
  \dot N_{\mathrm{acc}}(R_s) =  -  \int_{S^2_{r=R_s}}  \alpha c \sqrt{\gamma}\,J^r\,d\Omega.
  \label{eq:Noether_accretion_rate}
\end{equation}

\noindent This definition is obtained by integrating the continuity equation over the exterior region $r\ge R_s$. Since no physical outgoing flux can emerge from inside the event horizon, any inward-directed flux through $r=R_s$ is lost from the exterior domain. Therefore, $\dot N_{\mathrm{acc}}(R_s)$ provides a measure of the net particle-number accretion rate by the black hole \cite{BarrancoEtAl2011,SanchisGual2016Accreting}.

\subsection{Dimensionless rescaling and physical parameter space}

To perform the numerical integration of the KG equation in dimensionless units, we introduce the rescaled coordinates

\begin{equation}
  r=R_s\tilde r,  \qquad  t=\frac{R_s}{c}\tilde t,\nonumber
\end{equation}

\noindent where $r$ and $t$ are the radial and temporal coordinates in physical units, and $\tilde r$ and $\tilde t$ are their corresponding dimensionless computational counterparts. Substituting into the KG equation (\ref{eq:KG_covariant}) and multiplying by $R_s^2$, we obtain

\begin{equation}
  \widetilde{\Box}\phi-\tilde\mu^2\phi=0,\label{eq:KG_tildes}
\end{equation}

\noindent with the dimensionless mass parameter 

\begin{equation}
  \tilde\mu=\frac{R_s}{\lambdabar_C}.\nonumber
\end{equation}

\noindent This parameter coincides with the gravitational fine-structure constant $\alpha_G$ defined in Ref.~\cite{Alcubierre_2025}. Once the mass of the black hole and the mass of the boson are specified, $\tilde\mu$ becomes the only free parameter in the dimensionless formulation.

In the remainder of the paper, all evolution equations are written in the dimensionless variables $(\tilde t,\tilde r)$, although we drop the tildes for notational simplicity. Factors of $R_s$ and $c$ are restored only in figure labels and captions.

As an astrophysical reference scale of our analysis, we consider the supermassive black hole hosted at the center of the galaxy M87, with an estimated mass of about $6.5\times 10^{9}\,M_\odot$ according to the Event Horizon Telescope results \cite{EHT2019}. Throughout this work, we consider the boson mass within the range usually discussed in the context of fuzzy dark matter models, $m_b \sim 10^{-23}-10^{-19}\,\mathrm{eV}/c^2$ \cite{Ir_i__2017,Armengaud_2017,Hlozek_2015,10.1093/mnras/stv624,Sarkar_2016,Rindler_Daller_2012,Robles_2012}. As a reference scaling for the fixed black-hole mass of M87, taking $m_b=10^{-22}\,\mathrm{eV}/c^2$ yields $\tilde\mu\approx 9.7\times 10^{-3}$.

\section{First-order formulation and multipolar decomposition}
\label{sec:modal_reduction}

\subsection{First-order evolution system}

Using the conventions of the $3+1$ decomposition described, the covariant KG equation (\ref{eq:KG_tildes}) can be written as a system of first order equations in space and time by defining the auxiliary variables

\begin{equation}
  \psi_i = D_i \phi,
  \qquad
  \pi = n^\mu \nabla_\mu \phi,
\end{equation}

\noindent where $\psi_i$ represents the spatial gradient of the scalar field and $\pi$ its canonical momentum as measured by Eulerian observers. In terms of these variables, the KG equation takes the form

\begin{eqnarray}
  \partial_t \phi &=& \alpha \pi + \beta^i \psi_i, \nonumber \\
  \partial_t \psi_i &=& \beta^j D_j \psi_i + \psi_j D_i \beta^j + D_i(\alpha \pi), \nonumber\\
  \partial_t \pi &=& \beta^i D_i \pi + \alpha \left( D_i \psi^i + K\pi - \mu^2 \phi \right) + \psi^i D_i \alpha, \nonumber
\end{eqnarray}

\noindent where $K=-\nabla_\mu n^\mu$ is the trace of the extrinsic curvature of $\Sigma_t$. This system is equivalent to the covariant KG equation within the $3+1$ decomposition adapted to the background geometry.

\subsection{Multipolar decomposition of the wave packet}

For a spherically symmetric space-time, the KG operator
$\tilde{\Box}-\tilde{\mu}^2$, inherits the symmetry of the background and therefore commutes with the angular Laplacian on the two-sphere. As a consequence, the spherical harmonics form a natural angular basis that separates the angular dependence of the scalar field. This property allows to expand the field in spherical harmonics:

\begin{equation}
  \phi(t,r,\theta,\varphi)  =  \sum_{\ell=0}^{\infty}\sum_{m=-\ell}^{\ell}  \phi_{\ell m}(t,r)\,Y_{\ell m}(\theta,\varphi),
  \label{eq:phi_expansion}
\end{equation}

\noindent and the momentum variable $\pi$ is expanded in the same basis with modal coefficients $\pi_{\ell m}(t,r)$. In the first-order system, $\psi_i$ denotes the full spatial gradient of the field, however, after the angular decomposition, the reduced system only requires the radial modal component, which we denote by $\psi_{\ell m}(t,r) \equiv \partial_r \phi_{\ell m}(t,r)$, while the angular derivatives are encoded through the spherical harmonic eigenvalue $\ell (\ell + 1)$. Then, with this decomposition, each multipolar component evolves independently. For the Schwarzschild background, the equations for each mode $(\ell,m)$, that we call {\it reduced equations}, take the form

\begin{eqnarray}
  \partial_t \phi_{\ell m} &=& \alpha \pi_{\ell m} + \beta^r \psi_{\ell m}, \nonumber \\
  \partial_t \psi_{\ell m} &=& \partial_r \left(\alpha \pi_{\ell m} + \beta^r 
  \psi_{\ell m}\right), \label{eq:system_lm} \\
  \partial_t \pi_{\ell m}  &=&  \frac{1}{r^2\sqrt{\gamma_{rr}}}  \partial_r\!\left(\alpha r^2\sqrt{\gamma_{rr}}\,\gamma^{rr}\psi_{\ell m}\right)  +  \beta^r \partial_r \pi_{\ell m}
  \nonumber \\
  &+&
  \alpha K \pi_{\ell m}  -  \alpha\left(\tilde{\mu}^2+\frac{\ell(\ell+1)}{r^2}\right)\phi_{\ell m}. \nonumber
\end{eqnarray}

\noindent The original three-dimensional problem reduces to a collection of independent $1+1$ dimensional evolution systems, one for each multipole. The angular dependence is encoded in the harmonic labels $(\ell,m)$, while the effect of the angular sector appears explicitly through the term $\ell(\ell+1)/r^2$.

For a fixed $\ell$, Eq.~\eqref{eq:system_lm} does not have explicit dependence on $m$. Therefore, modes with the same $\ell$ and identical radial initial data share the same radial dynamics, the same modal Noether balance inherited from Eq.~\eqref{eq:continuity}, and the same accretion diagnostic inherited from Eq.~\eqref{eq:Noether_accretion_rate}. Under these conditions, the value $m$ specifies the linearly independent angular component associated with the common radial solution.

\subsection{Multipolar decomposition of conserved quantities}

The decomposition into spherical harmonics is inherited to the total Noether charge and the accretion rate Eqs.~\eqref{eq:Noether_charge} and \eqref{eq:Noether_accretion_rate}, which decompose as sums over multipoles:

\begin{equation}
  N(t)=\sum_{\ell=0}^{\infty}\sum_{m=-\ell}^{\ell} N_{\ell m}(t),\nonumber
\end{equation}
\begin{equation}
  \dot N_{\mathrm{acc}}(t)=\sum_{\ell=0}^{\infty}\sum_{m=-\ell}^{\ell}\dot N_{\ell m,\mathrm{acc}}(t).\nonumber
\end{equation}

\noindent Both scalars, $N_{\ell m}(t)$ and $\dot N_{\ell m,\mathrm{acc}}(t)$ are obtained from the corresponding mode contributions to the Noether current. Since each mode satisfies an independent evolution equation, each mode also satisfies its own continuity equation and therefore carries an individually conserved particle-number balance. The global conservation law is recovered by summation over all multipoles.

The accretion process can therefore be analyzed independently in each angular mode, while the total charge and the total accretion rate are recovered as sums over individual mode contributions.

\subsection{Mode-by-mode representation}

Once the angular dependence has been separated, each mode coefficient $\phi_{\ell m}(t,r)$ is a complex function of $t$ and $r$, and can therefore be written in amplitude-phase form as

\begin{equation}
  \phi_{\ell m}(t,r)  =  \sqrt{\varrho_{\ell m}(t,r)}\,e^{iS_{\ell m}(t,r) / \hbar},\nonumber
\end{equation}

\noindent where $\varrho_{\ell m}\ge 0$ is the squared amplitude of the mode and $S_{\ell m}$ its phase. For each mode, we define

\begin{equation}
  \omega_{\ell m}\equiv -\frac{1}{\hbar}\,n^\mu\nabla_\mu S_{\ell m},  \qquad  k_{\ell m}\equiv \frac{1}{\hbar} \partial_r S_{\ell m},\nonumber
\end{equation}

\noindent which are the counterparts of the frequency and spatial wavenumber defined in Eq.~\eqref{eq:phase_frequency_wavenumber}. Thus, $\omega_{\ell m}$ is the mode frequency measured by Eulerian observers, and $k_{\ell m}$ is the radial mode wavenumber.

The substitution of the Madelung form into the reduced equations yields the modal counterparts of the continuity equation \eqref{eq:Madelung_continuity_covariant} and the phase equation \eqref{eq:Madelung_HJ_covariant}. The imaginary part gives the continuity equation associated with the modal Noether current,

\begin{equation}
  \partial_t (\sqrt{\gamma}\,\rho_{n,\ell m})  +  \partial_r(\alpha\sqrt{\gamma}\,J^r_{\ell m})
  =  0,\nonumber 
\end{equation}

\noindent where

\begin{equation}
  \rho_{n,\ell m}  =  \varrho_{\ell m}\omega_{\ell m},  \qquad
  J^r_{\ell m}  =  \varrho_{\ell m}  \left(    \gamma^{rr}k_{\ell m}    -    \frac{\beta^r}{\alpha }\omega_{\ell m}  \right).\nonumber
\end{equation}

\noindent The real part yields the Hamilton--Jacobi type equation for the modal phase,

\begin{equation}
  -\omega_{\ell m}^2  +  \gamma^{rr}k_{\ell m}^2  +  \bar{\mu}^2  +  \frac{\ell(\ell+1)}{r^2}  -  Q_{\ell m}  =  0,
  \label{eq:Madelung_HJ_lm}
\end{equation}

\noindent where $Q_{\ell m}$ denotes the modal quantum potential associated with the
radial-temporal part of the d'Alambertian after the angular dependence has
been separated.

In this representation, the evolution of each mode is governed by three contributions: the mass term, the angular barrier $\ell(\ell+1)/r^2$, and the modal quantum potential $Q_{\ell m}$, which depends on the local structure of the squared field amplitude $\varrho_{\ell m}$.

\section{Numerical methods}
\label{sec:numerical_approach}

In practice, the multipolar expansion is truncated at a finite maximum value $\ell_{\max}$. After the spherical-harmonic decomposition, the truncation contains $(\ell_{\max}+1)^2$ independent $1+1$-dimensional evolution problems of type (\ref{eq:system_lm}), one for each mode $(\ell,m)$. Since the reduced equations  \eqref{eq:system_lm} are independent of $m$, the radial evolution and modal accretion diagnostic are identical for all modes with fixed $\ell$ and identical radial initial data, the production runs of our parameter-space exploration are evolved only for one value of $m$ for each $\ell$, chosen as $m=0$ for convenience.

To emphasize the numerical advantage of this representation, consider an equivalent three-dimensional angular-grid description of a field with multipolar content up to $\ell_{\max}$. Such a grid would have to satisfy the spherical Nyquist sampling criterion for the angular band limit $\ell_{\max}+1$, with typical requirements $N_\theta\sim \ell_{\max}+1$ and $N_\varphi\sim 2\ell_{\max}+1$ \cite{sph_FT}. For $\ell_{\max}=80$, this corresponds to approximately $81\times161$ angular points at each radial position. In contrast, the multipolar formulation treats the angular dependence analytically and replaces the angular-grid evolution by independent radial problems for the selected modes.

In our approach, the angular dependence is treated analytically through the spherical-harmonic basis, and the numerical evolution is performed independently for each representative multipolar component. In this way, each reduced system \eqref{eq:system_lm} is solved as a 1+1 initial value problem along the radial domain $r\in [r_{\rm min},r_{\rm max}]$, uniformly discretized with resolution $\Delta r= (r_{\rm max} - r_{\rm min})/N_r$, with $N_r=18000$ cells. We implement the excision method \cite{excision}, by setting the inner radius to $r_{\rm min}=0.5\,R_s$, located inside the event horizon. 

The outer boundary is artificial, since the physical exterior domain extends to spatial infinity. Although an outgoing characteristic condition is imposed there, this condition is only a local approximation to a transparent boundary for the scalar field. We place this boundary at $r_{\max}=600\,R_s$, in order to ensure that potential residual boundary effects, such as a spurious reflected components or artificial loss of charge through the outer edge of the grid, are causally disconnected from the near-horizon region where the accretion diagnostics is evaluated during the time domain considered here, which is set to $t\in[0,T]$, with $T=200 R_s/c$.

The reduced evolution system for $\phi_{\ell m}$, $\psi_{\ell m}$, and $\pi_{\ell m}$ is strongly hyperbolic, and its principal part has characteristic speeds

\begin{equation}
  \lambda_\pm=\pm \alpha\sqrt{\gamma^{rr}}-\beta^r,\nonumber
\end{equation}

\noindent associated with characteristic fields

\begin{equation}
  e_\pm=\pi_{\ell m}\mp \sqrt{\gamma^{rr}}\,\psi_{\ell m}.\nonumber
\end{equation}

\noindent These fields describe the inward-outward radial propagation directions of the system and provide the characteristic variables for imposing boundary conditions.

At the outer boundary, the inward characteristic field $e_-$ is set to zero. At the inner boundary, the outward characteristic field $e_+$ is suppressed. Since each characteristic condition determines only one linear combination of $\pi_{\ell m}$ and $\psi_{\ell m}$, the fields $\phi_{\ell m}$ and $\psi_{\ell m}$ are obtained in ghost zones by third-order Lagrange extrapolation from points within the numerical domain, and $\pi_{\ell m}$ is then reconstructed from the corresponding characteristic relation. Specifically, at the outer boundary the condition $e_-=0$ implies

\begin{equation}
  \pi_{\ell m}=-\sqrt{\gamma^{rr}}\,\psi_{\ell m},\nonumber
\end{equation}

\noindent whereas at the inner boundary the condition $e_+=0$ gives

\begin{equation}
  \pi_{\ell m}=+\sqrt{\gamma^{rr}}\,\psi_{\ell m}.\nonumber
\end{equation}

\noindent This method preserves a smooth extension of the fields into the ghost zones we use within the excision boundary, consistently with the causal structure of the space-time inside the horizon.

Radial derivatives are approximated with second-order finite differences on the uniform grid, and the advection terms of the type $\beta^r\partial_r$, use causally connected  unbalanced stencils \cite{alcubierre1994investigations}. The resulting semi-discrete equations are integrated in time with a third-order Runge-Kutta scheme. Since the characteristic speeds satisfy $|\lambda_\pm|\leq 1$ and depend only on the fixed background geometry, all simulations are performed with a constant CFL factor $C=0.25$.

The numerical consistency and self-convergence of these methods is detailed in Appendix~\ref{app:numerical_validation}. The validation consists of three tests. First, a Richardson self-convergence check of the Noether charge verifies the expected second-order convergence. Second, a global Noether-charge calculation shows that the conservation defect associated with the flux-balance law converges to zero when increasing resolution. Third, an inner-boundary-location test confirms that moving the excision radius within the horizon has no measurable effect on the exterior evolution, which shows that inner boundary conditions work, consistently with the causal structure of the space-time. These tests support the numerical methods used in the set of production runs of our analysis described below.

\section{Black hole response to wave packets}
\label{sec:mono_waves}

\subsection{Nearly monochromatic initial data}

As follows from the mode reduced system \eqref{eq:system_lm}, the radial dynamics
of each multipolar component is determined by the index $\ell$, while the
index $m$ only selects the angular harmonic used in the reconstruction of the
field. For fixed $\ell$ and identical radial initial data, choosing any value
of $m$ gives the same radial evolution and the same modal accretion fraction.
In our analysis we therefore choose $m=0$ arbitrarily for each $\ell$. This
choice is made only to obtain a simple axisymmetric representation of the
general field morphology. We denote the corresponding variables by
$\phi_\ell \equiv \phi_{\ell 0}$, $\psi_\ell \equiv \psi_{\ell 0}$, and
$\pi_\ell \equiv \pi_{\ell 0}$. The scalar fields shown below are 
axisymmetric reconstructions of the representative sectors

\begin{equation}
  \phi(t,r,\theta,\varphi)  =  \sum_{\ell=0}^{\ell_{\max}}  \phi_\ell(t,r)\,Y_{\ell 0}(\theta,\varphi).
  \label{eq:axisymmetric_reconstruction}
\end{equation}

\noindent The accretion efficiency obtained from these representative sectors apply to
any $m$ with the same $\ell$ and the same radial initial profile.

For each $\ell$, the initial scalar field profile is prescribed as a radially localized Gaussian packet modulating a monochromatic carrier phase $k_0$:

\begin{equation}
  \phi_{\ell}(t=0,r)=  \exp\!\left[-\frac{(r-r_0)^2}{2\sigma^2}\right]e^{-ik_0 r},
  \label{eq:initial_phi}
\end{equation}

\noindent where $r_0$ and $\sigma$ are the central radius and width of the Gaussian distribution. In Madelung variables, $\phi_\ell=\sqrt{\varrho_\ell}\,e^{iS_\ell/\hbar}$, which corresponds initially
to

\begin{equation}
 \sqrt{ \varrho_\ell}(t=0,r)=  \exp\!\left[-\frac{(r-r_0)^2}{\sigma^2}\right],
  ~~  S_\ell(t=0,r)=-\hbar k_0 r .\nonumber
\end{equation}

\noindent Therefore, the local radial wavenumber at initial time is

\begin{equation}
  k_\ell(t=0,r)=\frac{1}{\hbar}\partial_r S_\ell(t=0,r)=-k_0 .
\nonumber 
\end{equation}

\noindent Thus, $k_0$ is the imposed local radial wavenumber of the initial packet and the localization is entirely contained in the Gaussian envelope. The auxiliary fields are initialized as

\begin{equation}
  \psi_\ell(t=0,r)=\partial_r\phi_\ell(t=0,r),
  \label{eq:initial_psi}
\end{equation}

\noindent and

\begin{equation}
  \pi_\ell(t=0,r)=\sqrt{\gamma^{rr}(r)}\,\psi_\ell(t=0,r).
  \label{eq:initial_pi}
\end{equation}

\noindent This choice selects the ingoing modes of the first-order fields at
initial time. We fix the overall phase so that the associated Noether charge
$N_\ell$ is positive, where $N_\ell>0$ is interpreted as particle,
rather than antiparticle, content. The overall amplitude fixes the initial
Noether charge of each simulation. Also notice that since the KG equation is linear in the test-field regime, the accreted fractions reported below are invariant under a global rescaling of the field amplitude.

For the ingoing part imposed by Eqs.~\eqref{eq:initial_psi} and
\eqref{eq:initial_pi}, evaluating the Hamilton--Jacobi equation for each mode,
Eq.~\eqref{eq:Madelung_HJ_lm}, at initial time gives

\begin{equation}
  Q_\ell(t=0,r)  =  \tilde{\mu}^2+\frac{\ell(\ell+1)}{r^2}.
\nonumber
\end{equation}

\noindent This relation shows that ingoing data are initially compatible with the modal Hamilton--Jacobi balance equation. Finally, over the core of the packet, the angular contribution is suppressed by the large initial radius $r_0\gg R_s$.

\subsection{Example of a wave-packet evolution}
\label{subsec:representative_evolution}

All wave-packet simulations in this work use
$r_0 = 80\,R_s$ and $\sigma = 10\,R_s$. These values place the packet
sufficiently far from the black hole to lie in an approximately Minkowskian
region while keeping a localized profile inside the computational domain. The
evolutions are carried out up to $T = 200\,R_s/c$, which is long enough to
capture the accretion stage and short enough to avoid potential contamination
from outer-boundary reflections at $r_{\max}=600\,R_s$ due to causal disconnection.

For illustration of our analysis, we first discuss a representative configuration with $m_b=10^{-22}\,\mathrm{eV}/c^2$ and central radial wavenumber
$k_0=1.5R_s^{-1}$. 
From the evolved set, we focus on the representative sectors $\ell=3,4,5$, which illustrate efficient accretion, partial accretion, and scattering for a fixed $k_0$.
These sectors share the same initial radial carrier and Gaussian envelope, while their
different angular barriers produce different interactions with the black-hole
potential.

In order to quantify the response to each mode, we use the accumulated accreted fraction

\begin{equation}
  \eta_{\ell,\mathrm{acc}}(t)  =  \frac{1}{N_\ell(0)}  \int_0^t  \dot{N}_{\ell,\mathrm{acc}}(t')\,dt' .
  \label{eq:eta_l_acc_time}
\end{equation}

\begin{figure}[t]
  \centering
  \includegraphics[width=\columnwidth]{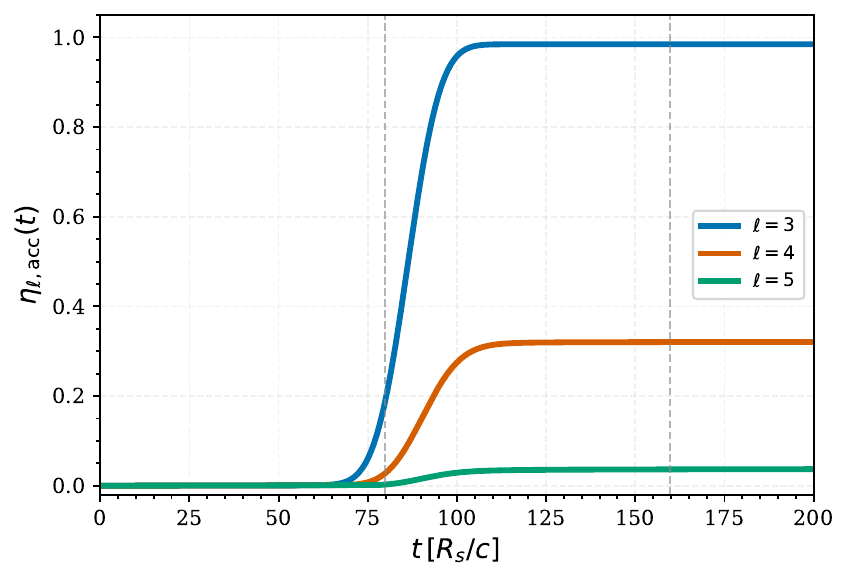}
  \caption{
  Accumulated accreted fraction $\eta_{\ell,\mathrm{acc}}(t)$ for the
  representative sectors $\ell=3,4,5$ in the case
  $m_b=10^{-22}\,\mathrm{eV}/c^2$ and $k_0=1.5R_s^{-1}$. The final
  values identify the $\ell=3$ sector as efficiently accreted, the $\ell=4$
  sector as partially accreted, and the $\ell=5$ sector as mostly scattered. The vertical dotted lines mark the times used later in Fig.~\ref{fig:radial_profiles_rep}.   }
  \label{fig:eta_rep}
\end{figure}

\noindent With this convention, $\eta_{\ell,\mathrm{acc}}=0$ corresponds to no net
accretion of the initial Noether charge of the mode $\ell$, while
$\eta_{\ell,\mathrm{acc}}=1$ corresponds to complete accretion. The time
evolution of $\eta_{\ell,\mathrm{acc}}(t)$ is shown in
Fig. \ref{fig:eta_rep}. The accreted fraction grows mainly during the interval
in which the packet reaches the near-horizon region and then approaches a
saturation value. The final values classify the three selected responses: the
$\ell=3$ sector is efficiently accreted, the $\ell=4$ sector is partially
accreted, and the $\ell=5$ sector is dominated by exterior scattering.

\begin{figure*}[t]
  \centering
  \includegraphics[width=\textwidth]{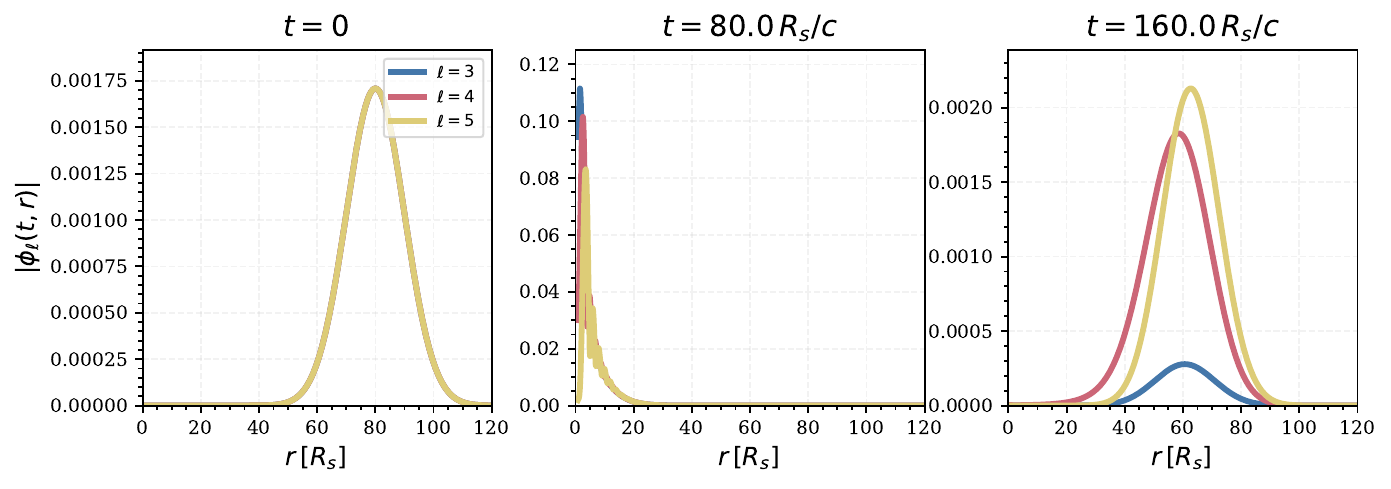}
  \caption{
  Radial profiles of $|\phi_\ell(t,r)|$ for the representative case
  $m_b=10^{-22}\,\mathrm{eV}/c^2$ and $k_0=1.5R_s^{-1}$.
  The selected sectors $\ell=3,4,5$ are shown at $t=0$, $t=80R_s/c$, and
  $t=160R_s/c$. The initially equal density shells evolve into different
  late-time exterior remnants: $\ell=3$ leaves the smallest remnant after
  efficient accretion, $\ell=4$ leaves an intermediate component, and
  $\ell=5$ shows the largest scattered profile.
  }
  \label{fig:radial_profiles_rep}
\end{figure*}

In Figure \ref{fig:radial_profiles_rep} we show the corresponding radial profiles $|\phi_\ell(t,r)|$ at $t=0$, $t=80R_s/c$, and $t=160R_s/c$. At initial time, the selected modes have the same radial envelope centered at $r_0=80R_s$. Around $t\simeq80R_s/c$, the packet is concentrated close
to the black hole, precisely during the time interval in which
$\eta_{\ell,\mathrm{acc}}$ grows. By $t=160R_s/c$, the profiles have
differences related to their accretion efficiencies: the more efficiently
accreted $\ell=3$ sector has the smallest exterior remnant, the partially
accreted $\ell=4$ sector shows an intermediate component, and the mostly
scattered $\ell=5$ sector contains the largest exterior profile. The late-time
radial hierarchy is therefore the spatial counterpart of the Noether-charge
hierarchy measured by $\eta_{\ell,\mathrm{acc}}$.

We also track the radial phase content of the exterior field through a
geometrically weighted average of the local wavenumber. Since the spatial
volume element on a time slice is $\sqrt{\gamma}\,dr\,d\Omega$, the radial
weight after angular integration is $r^2\sqrt{\gamma_{rr}}$, we define

\begin{equation}
\langle k_\ell\rangle(t)=\frac{\int_{R_s}^{r_{\max}} k_\ell(t,r)|\phi_\ell(t,r)|^2
r^2\sqrt{\gamma_{rr}}\,dr}{\int_{R_s}^{r_{\max}} |\phi_\ell(t,r)|^2r^2\sqrt{\gamma_{rr}}\,dr}.
\nonumber 
\end{equation}

\begin{figure}[b]
  \centering
  \includegraphics[width=\columnwidth]{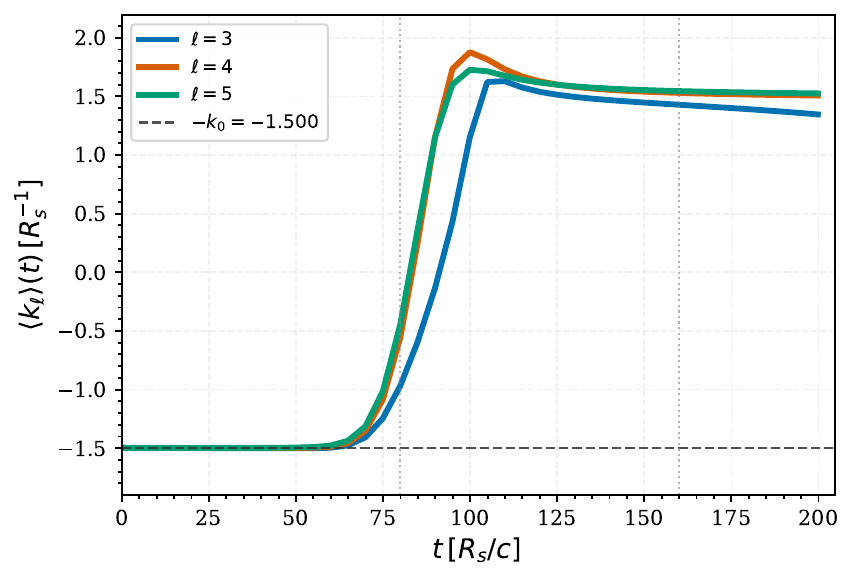}
  \caption{
  Geometrically weighted radial average $\langle k_\ell\rangle(t)$ for the
  representative sectors $\ell=3,4,5$. The dashed horizontal line marks the
  initial value $-k_0$, imposed by the phase factor $e^{-ik_0r}$. The sign
  change indicates that the exterior support becomes dominated by an outgoing
  scattered component after the interaction with the black hole. The late-time
  values characterize the radial-wavenumber content of the surviving exterior
  remnant.
  }
  \label{fig:kmean_rep}
\end{figure}

\noindent In Figure \ref{fig:kmean_rep} we show the selected sectors $\ell=3,4,5$, which  initially satisfy $\langle k_\ell\rangle\simeq -k_0$, consistent with the initial phase
$e^{-ik_0r}$. During the interaction stage, $\langle k_\ell\rangle$ changes
sign and becomes positive. This change of sign indicates that the exterior
field is dominated by an outgoing scattered component.

The post-interaction value of $\langle k_\ell\rangle$ also contains information
about the radial content of the surviving exterior packet. The initial data are
nearly monochromatic, but the Gaussian envelope gives the packet a finite
spectral width around the carrier scale $k_0$. The black-hole potential acts
as a mode-dependent filter, since the portions of the packet that are transmitted
toward the horizon and those scattered back to the exterior do not sample the
same radial-wavenumber content. Consequently, the late-time value of
$\langle k_\ell\rangle$ should be interpreted as the mean radial wavenumber of
the outgoing remnant, not as the wavenumber of the original packet as a whole.
This effect is most visible for $\ell=3$, where the dominant ingoing component
is efficiently accreted and the remaining exterior field is small. The decrease
in $|\langle k_3\rangle|$ after its maximum indicates that the surviving
outgoing remnant has a radial-wavenumber content shifted to smaller absolute
values relative to its initial value.

\begin{figure*}[t]
  \centering
  \includegraphics[width=\textwidth]{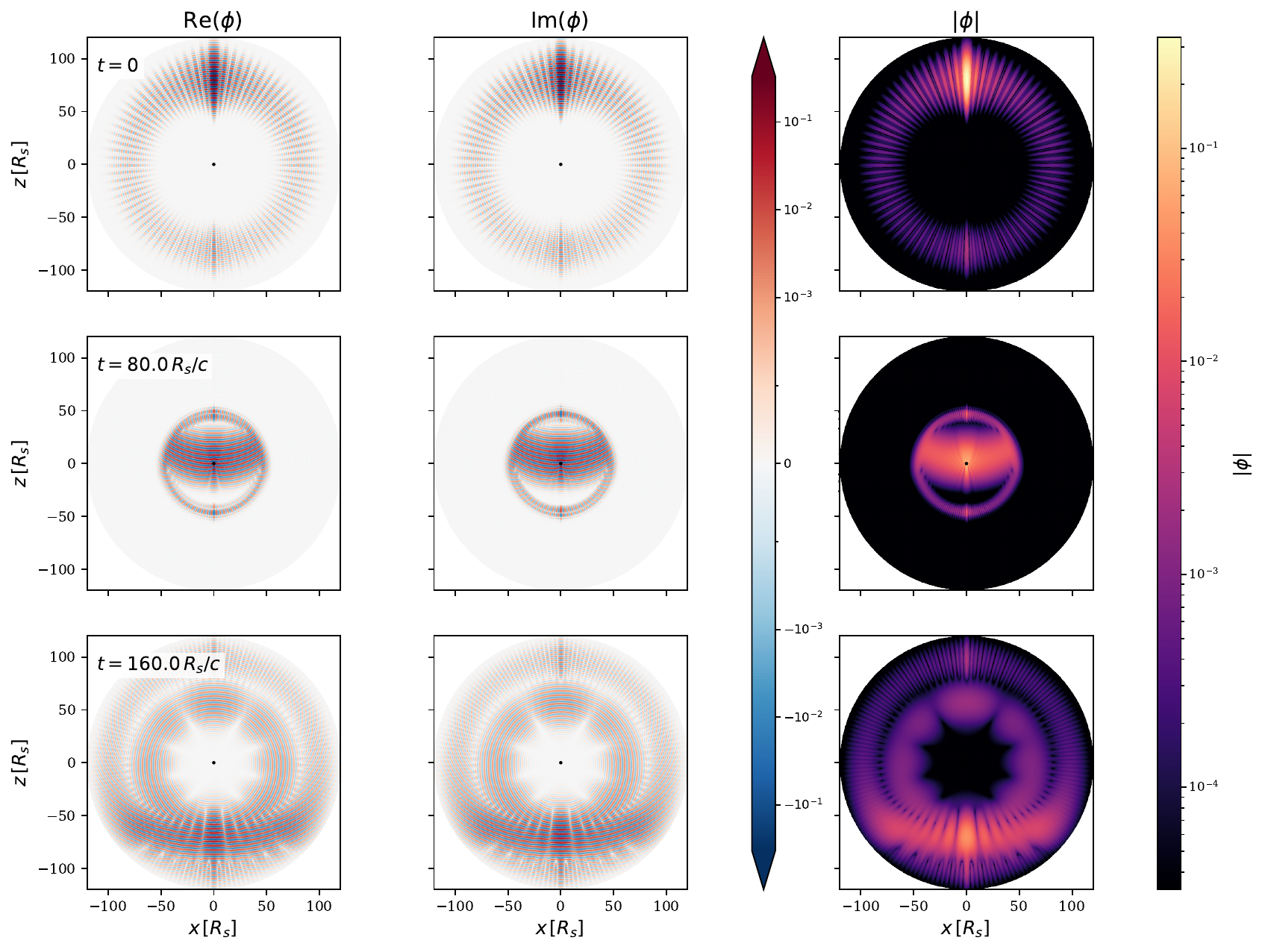}
  \caption{
  Reconstructed axisymmetric scalar field from the solutions for $m=0$, with $\ell_{\rm max}=80$.
  Columns show $\mathrm{Re}(\phi)$, $\mathrm{Im}(\phi)$, and $|\phi|$ projected on the Cartesian   $xz-$plane, while rows correspond to $t=0$, $t=80R_s/c$, and
  $t=160R_s/c$. 
  The late-time panels show the angular nodal structure of the exterior
  remnant after the smoother, more efficiently accreted lower-$\ell$
  contribution has been accreted.
  }
  \label{fig:reconstructed_rep}
\end{figure*}

Finally, the mode-dependent accretion also changes the spatial morphology of the
reconstructed axisymmetric scalar field. Figure \ref{fig:reconstructed_rep}
shows $\mathrm{Re}(\phi)$, $\mathrm{Im}(\phi)$, and $|\phi|$ in the Cartesian 
$xz-$plane, obtained from the truncated reconstruction Eq.~\eqref{eq:axisymmetric_reconstruction}.
For the representative configuration, this reconstruction uses the full set of
evolved sectors included in the truncation up to $\ell_{max}=80$, not only the three sectors
displayed in Figs. \ref{fig:eta_rep}, \ref{fig:radial_profiles_rep} and \ref{fig:kmean_rep}. At $t=0$, the field is localized away from the
black hole and contains the angular structure imposed by the multipolar
superposition. At $t=80R_s/c$, the packet is strongly distorted as it
interacts with the near-horizon region. At $t=160R_s/c$, the remaining
exterior field displays a more high multipole pattern. This happens
because the lower-$\ell$ components are more efficiently accreted. On the other hand, the exterior remnant is more massive for the less absorbed higher-$\ell$
components. The
reconstruction therefore shows how selective modal absorption reshapes not only
the radial profiles, but also the angular structure of the field that remains
outside the black hole.

Together, these diagnostics show how the value of
$\eta_{\ell,\mathrm{acc}}$ is built dynamically. The initial packet propagates
inward, interacts with the near-horizon region, loses part of its Noether
charge through the horizon, and leaves an outgoing exterior remnant whose
radial and angular structure depend on the multipolar sector. This
representative simulation is therefore a time-domain realization of the
mode-dependent transition between absorption and scattering. In the next
subsection, the same charge-based diagnostic is used to map this transition
systematically in the $(k_0,\ell)$ parameter space.

\subsection{Accretion maps in parameter space}
\label{subsec:accretion_maps}

The representative evolution above, shows how different multipolar sectors can correspond to efficient accretion, partial accretion, or exterior scattering for a given initial radial wavenumber. We now extend such analysis to a systematic survey in the $(k_0,\ell)$ plane and a given set of values of the bsoson mass: 

\begin{eqnarray}    
  m_b =&
  0&,\,
  10^{-23},\,
  10^{-22},\,
  10^{-21},\,
  2\times10^{-20},\, 
  4\times10^{-20},\,\nonumber \\
  &6&\times10^{-20},\,
  8\times10^{-20},\,
  10^{-19}\,
  \ {\rm eV}/c^2 .\label{eq:mbs}
\end{eqnarray}

\noindent For each of the boson mass values considered, we perform simulations with $100$ regularly spaced values of the central radial wavenumber in the interval

\begin{equation}
  k_0\in (0,10]\,R_s^{-1}.\nonumber
\end{equation}

\noindent For each pair $(k_0,\ell)$, the modal accreted fraction is evaluated at the
final time $T$ using Eq.~\eqref{eq:eta_l_acc_time}. Thus, each point in the
maps corresponds to a full time-domain evolution  encodes the final fraction of the initial Noether charge absorbed by
the black hole. By the $m$-degeneracy of the reduced equations, the same radial
response applies to any mode with the same $\ell$ and the same radial initial
scalar field profile.

The maps display a transition between reflection and accretion regions in the $(k_0,\ell)$ plane. To intepret this transition, we use the effective
potential associated with each multipolar component $\phi_{\ell m}$ in
Eq.~\eqref{eq:phi_expansion}. Introducing the rescaled field
$u_{\ell m}=r\phi_{\ell m}$ and the tortoise coordinate $r^*$, the radial KG
equation can be written in the one-dimensional wave-like form

\begin{equation}
  \partial_t^2 u_{\ell m}  -  \partial_{r^*}^2 u_{\ell m}  +  V_{\mathrm{eff}}(r;\ell,\tilde{\mu})u_{\ell m}  =  0,\nonumber
\end{equation}

\noindent with effective potential \cite{BarrancoEtAl2011}

\begin{equation}
  V_{\mathrm{eff}}(r;\ell,\tilde{\mu})  =  \left(1-\frac{R_s}{r}\right)
  \left(    \tilde{\mu}^2    +    \frac{\ell(\ell+1)}{r^2}    +    \frac{R_s}{r^3}
  \right).
  \nonumber 
\end{equation}

\noindent Since the initial packet is centered at a large radius, where the spacetime is approximately Minkowskian, we assume its carrier frequency satisfies the asymptotic dispersion relation

\begin{equation}
  \omega^2 \simeq k_0^2+\tilde{\mu}^2.\nonumber
\end{equation}

\noindent Defining

\begin{equation}
  V_{\max}(\ell,\tilde{\mu})  =  \max_{r>R_s} V_{\mathrm{eff}}(r;\ell,\tilde{\mu}),\nonumber
\end{equation}

\noindent the classical condition for transmission across the barrier is

\begin{equation}
  \omega^2 > V_{\max}.\nonumber
\end{equation}

\noindent Using the asymptotic dispersion relation, the corresponding threshold in the central radial wavenumber is

\begin{equation}
  k_0^{\mathrm{tr}}(\ell,\tilde{\mu})  =  \sqrt{V_{\max}(\ell,\tilde{\mu})-\tilde{\mu}^2}.
  \label{eq:k0tr_def}
\end{equation}

\noindent The threshold curve Eq.~\eqref{eq:k0tr_def} gives the classical reference for \fb{ the radial carrier scale required to cross the maximum of the effective barrier}. It is therefore used as a reference curve to identify the transition band between reflection-dominated and accretion-dominated regions in the maps.

\begin{figure}[tbp]
    \centering
    \includegraphics[width=\columnwidth]{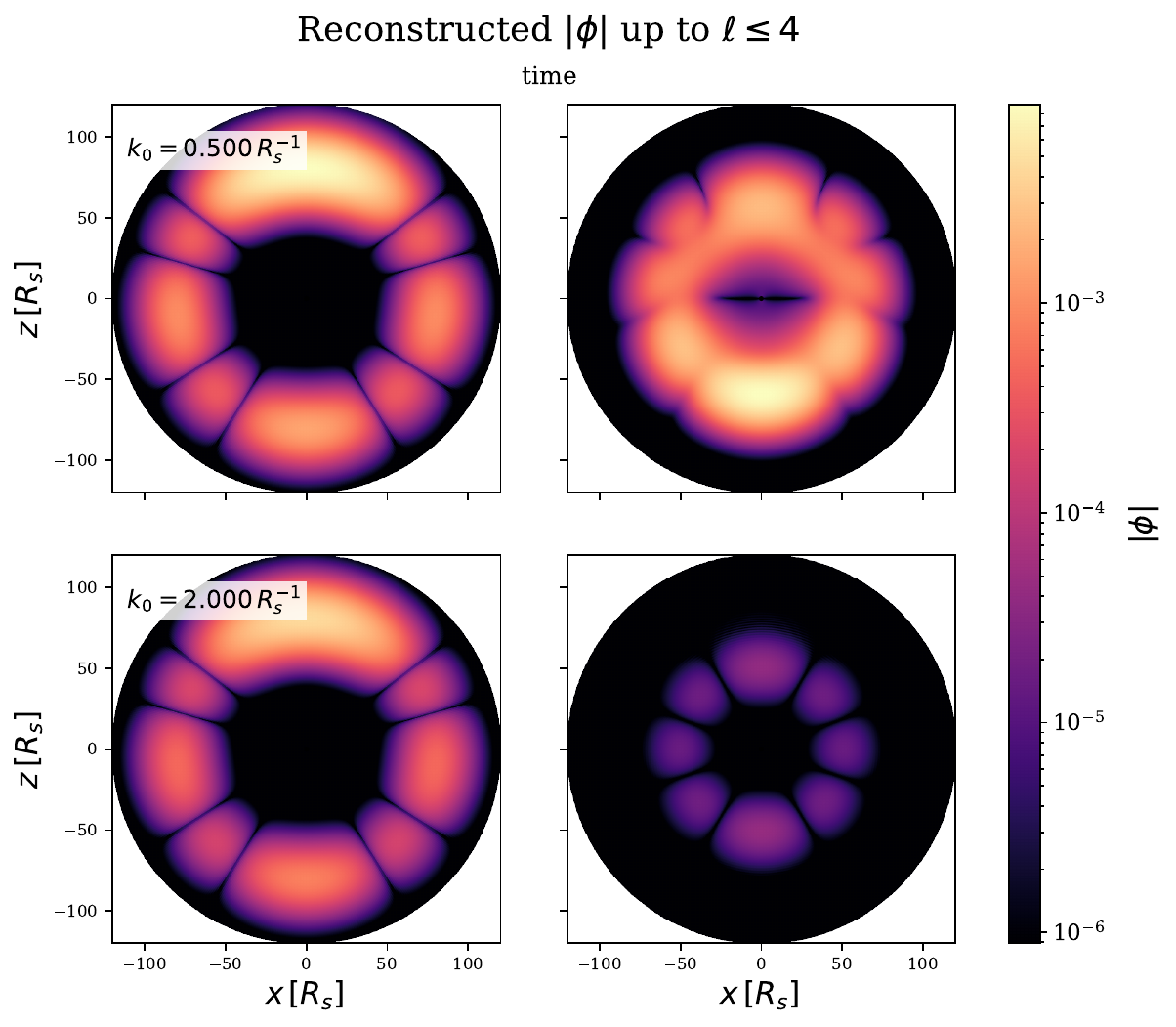}
    \caption{
    Reconstructed scalar field $|\phi|$ on the Cartesian $xz-$plane for
    two values of the initial radial wavenumber $k_0$. The reconstruction
    includes the representative $m=0$ sectors up to $\ell \leq 4$. Columns show the
    initial configuration at $t=0$ and the late-time configuration at
    $t=160R_s/c$. The upper row corresponds to
    $k_0=0.5\,R_s^{-1}$, which lies below the classical transmission threshold Eq.~\eqref{eq:k0tr_def}
    for several of the included modes, while the lower row corresponds to
    $k_0=2\,R_s^{-1}$, where more modes are above the classical transmission threshold.
    The late-time fields
    show how the map classification translates into exterior scattered
    structure or stronger depletion by accretion.
    }
    \label{fig:threshold_reconstructed_abs_phi}
\end{figure}

Before discussing the full parameter-space survey, Fig. \ref{fig:threshold_reconstructed_abs_phi}
shows how this map-based classification is reflected in the reconstructed
field. The figure compares two simulations with $m_b = 10^{-22}\,eV/c^2$ 
and values of $k_0 = 0.5,\,2.0\,R_s^{-1}$, using the reconstructed modulus
$|\phi|$ in the Cartesian $xz-$plane, including the representative $m=0$ sectors up to
$\ell \leq 4$. For this finite reconstruction, the accretion map indicates which
of the included multipoles are expected to be transmitted through the barrier
and which remain in the exterior. For $k_0=0.5R_s^{-1}$, several of the reconstructed components lie below or close to the transmission threshold Eq.~\eqref{eq:k0tr_def}, and the late-time field keeps an extended exterior remnant. For $k_0=2R_s^{-1}$, more of the included multipoles lie above of the transmission threshold, producing an efficient accretion of the exterior field at late times. Thus, the reconstructed angular patterns provide a direct spatial reading of the maps, the components classified as reflected or partially accreted determine the exterior nodal structure, whereas efficiently accreted components are depleted from the late-time field.

\begin{figure*}[t]
  \centering
  \includegraphics[width=0.80\textwidth]{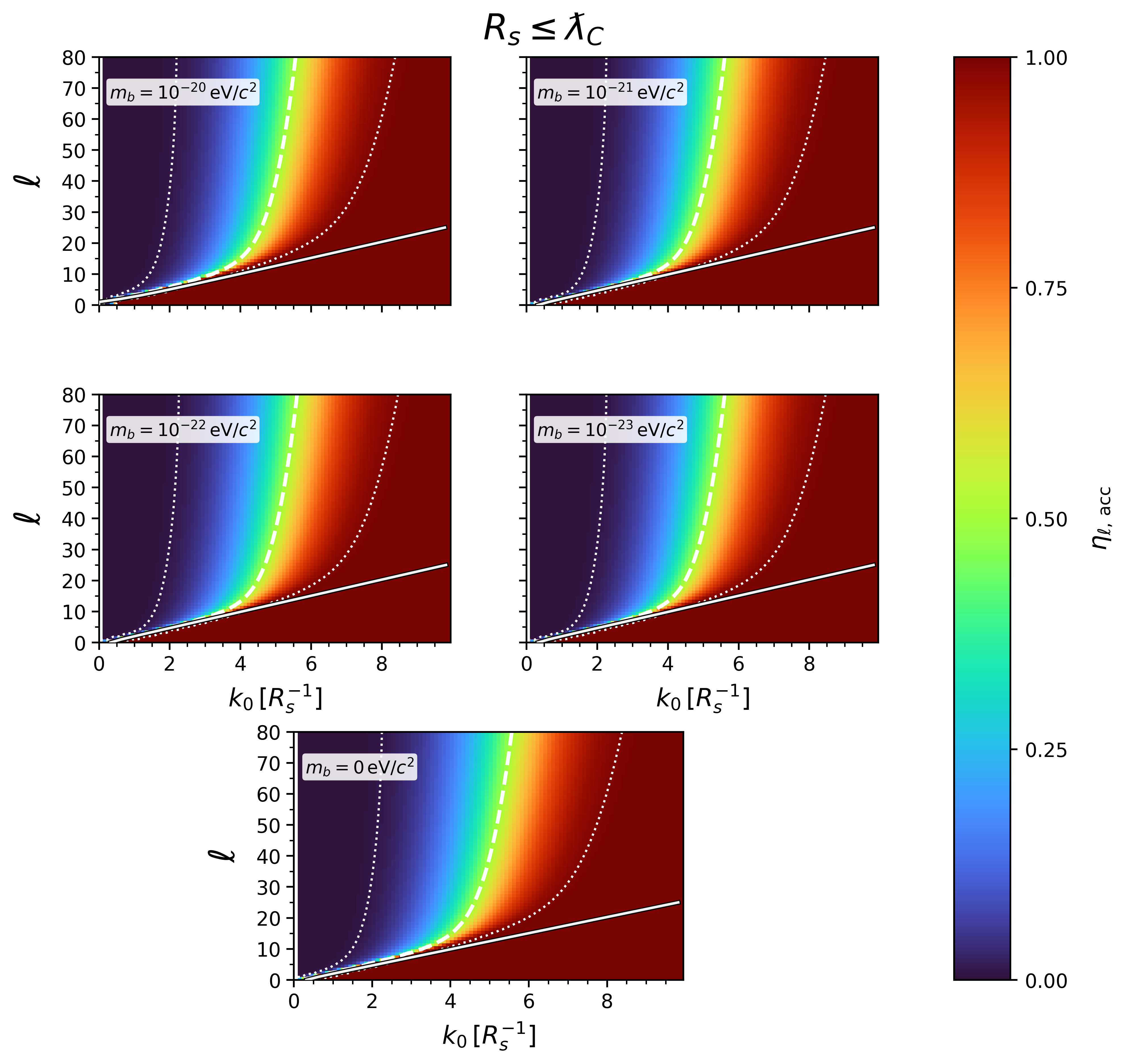}
  \caption{
  Accreted fraction $\eta_{\ell,\mathrm{acc}}(T)$ in the $(k_0,\ell)$ plane
  for boson masses satisfying $R_s\leq\lambdabar_C$. Colors encode
  $\eta_{\ell,\mathrm{acc}}(T)\in[0,1]$. The solid curve shows the classical
  transmission threshold $k_0^{\mathrm{tr}}(\ell)$ defined in
  Eq.~\eqref{eq:k0tr_def}. White dotted contours correspond to
  $\eta_{\ell,\mathrm{acc}}=0.01$ and $\eta_{\ell,\mathrm{acc}}=0.99$, and the
  white dashed contour corresponds to $\eta_{\ell,\mathrm{acc}}=0.5$. Notice that the values of $m_b$ do not include some of the values in (\ref{eq:mbs}) because they are forbidden by the relation between $R_s$ and $\lambdabar_C$.
  }
  \label{fig:Rs_lt_lambda}
\end{figure*}

\begin{figure*}[t]
  \centering
  \includegraphics[width=0.80\textwidth]{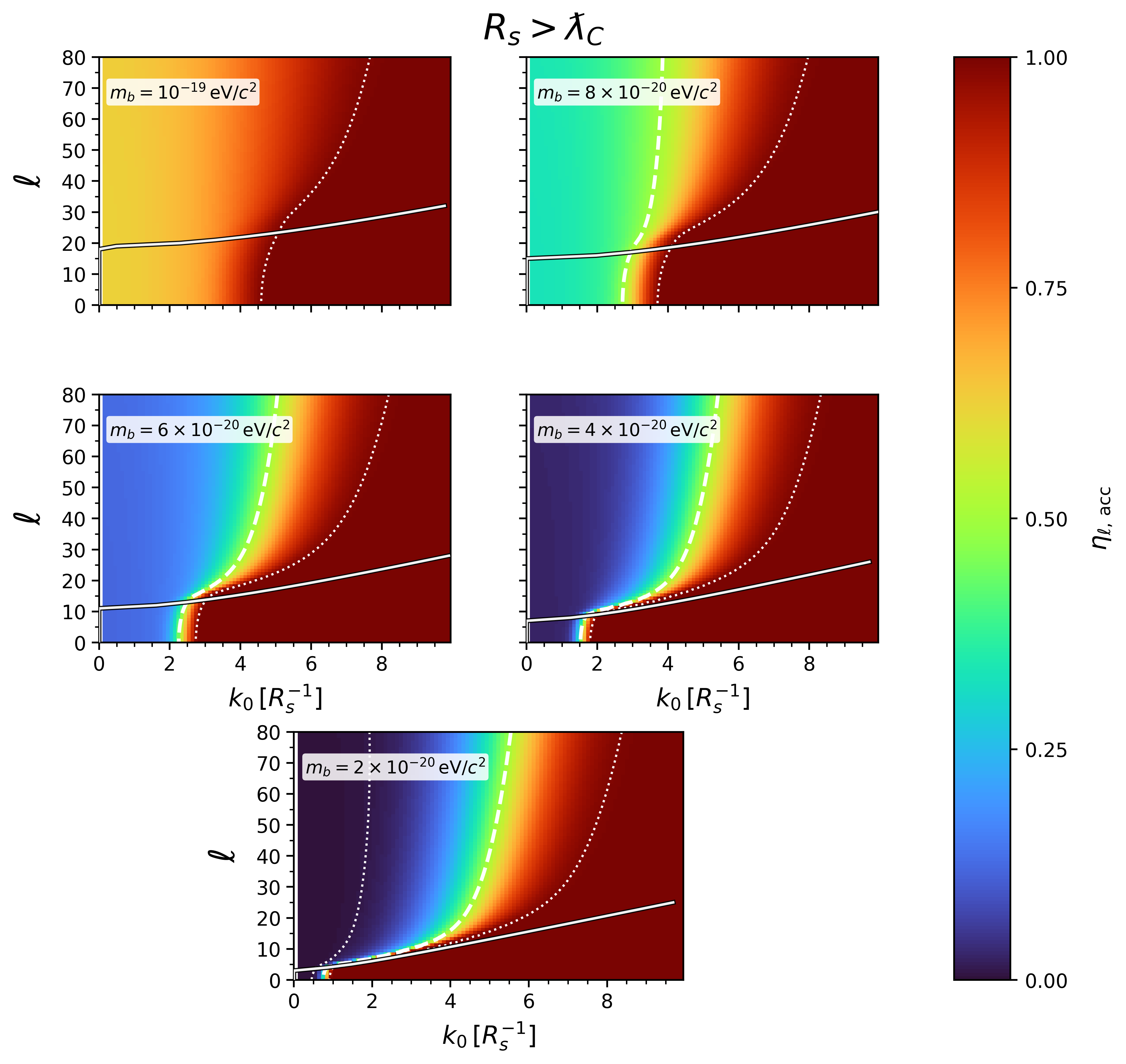}
  \caption{
  Accreted fraction $\eta_{\ell,\mathrm{acc}}(T)$ in the $(k_0,\ell)$ plane
  for boson masses in the regime $R_s>\lambdabar_C$. Colors encode
  $\eta_{\ell,\mathrm{acc}}(T)\in[0,1]$. The solid curve shows the classical
  transmission threshold $k_0^{\mathrm{tr}}(\ell)$ defined in
  Eq.~\eqref{eq:k0tr_def}. White dotted contours correspond to
  $\eta_{\ell,\mathrm{acc}}=0.01$ and $\eta_{\ell,\mathrm{acc}}=0.99$, and the
  white dashed contour corresponds to $\eta_{\ell,\mathrm{acc}}=0.5$.
  The low-$k_0$ region displays a partial-accretion floor that becomes more
  pronounced as the mass scale dominates the radial dynamics. Notice again, that the values of $m_b$ do not include some of the values in (\ref{eq:mbs}) because they are forbidden by the relation between $R_s$ and $\lambdabar_C$.
  }
  \label{fig:Rs_ge_lambda}
\end{figure*}

The resulting accretion maps are shown in Figs. \ref{fig:Rs_lt_lambda} and \ref{fig:Rs_ge_lambda}. In these maps, the color scale represents $\eta_{\ell,\mathrm{acc}}(T;k_0)\in[0,1]$. Values close to zero indicate that most of the modal charge remains in the exterior after the interaction, values close to unity indicate efficient accretion, and intermediate values correspond to partial accretion. In this
way, the maps provide a response diagram for the black hole, they identify
which multipolar sectors are transmitted through the effective barrier and
which sectors remain predominantly as exterior scattered components.

The exterior remnant selected by the accretion map can be viewed as a finite-time counterpat of  the long-lived massive-scalar configurations discussed in the context of scalar wigs, quasi-bound states, and quasi-stationary configurations \cite{Barranco_2011,Barranco2012ScalarWigs,Barranco2014SpectralWigs,SanchisGual2016}. In all of these cases, the same effective radial potential controls the competition between
exterior support and leakage through the horizon. \\

The results separate naturally according to the hierarchy between the Schwarzschild radius $R_s$ and the reduced Compton wavelength $\lambdabar_C$:

\begin{equation}
  R_s \leq \lambdabar_C,   \qquad  R_s > \lambdabar_C .\nonumber
\end{equation}

\underline{Regime $R_s \le \lambdabar_C$}. Figure \ref{fig:Rs_lt_lambda} shows the accretion maps for the case where the reduced Compton wavelength is comparable to or larger than the horizon scale, so that the mass term does not introduce a shorter intrinsic scale than the black-hole radius. The maps show a broad and curved transition between exterior retention and efficient accretion. As $\ell$ increases, the angular contribution
$\ell(\ell+1)/r^2$ raises the effective barrier, and the value of $k_0$
required for efficient accretion increases.

A relevant feature of this regime is that the maps have very similar structure for the different boson masses used. The transition remains broad and curved, with its location mainly controlled by the angular barrier. Therefore, for $R_s\leq\lambdabar_C$, the response is mainly controlled by the radial carrier scale and the angular contribution, whereas the boson mass produces only mild quantitative changes over the range considered. The classical threshold $k_0^{\mathrm{tr}}(\ell,\tilde{\mu})$ follows the main separation between the reflection-dominated and
accretion-dominated regions, which confirms that the effective barrier provides
the correct organizing structure for these maps.

\underline{Regime $R_s > \lambdabar_C$}. Figure \ref{fig:Rs_ge_lambda} shows the  maps for this regime, where the reduced Compton wavelength is shorter than the horizon scale, and the mass scale has a stronger effect on the radial transport of the packet. The transition in the $(k_0,\ell)$ plane becomes more compressed in $k_0$ over a wide range of multipoles. The response is therefore weakly dependient of $\ell$ at low-$k_0$ region, that is, many sectors move from partial accretion to efficient accretion within a narrower interval of the radial carrier wavenumber.

One very interesting new feature is the partial-accretion floor visible at low
$k_0$. For small boson masses, low-$\ell$ sectors with small $k_0$ can be
efficiently absorbed because the angular barrier is weak. As the boson mass
increases and $R_s/\lambdabar_C$ becomes larger than unity, these same low-$k_0$
and low-$\ell$ sectors no longer saturate immediately toward
$\eta_{\ell,\mathrm{acc}}\simeq1$. Instead, they enter a partial-accretion
regime, a significant fraction of the Noether charge crosses the horizon, and a
non-negligible remnant remains in the exterior at the final time.

This behavior follows from the radial phase content of the initial data. The
asymptotic dispersion relation gives $\omega^2\simeq k_0^2+\tilde{\mu}^2$. For fixed small $k_0$, increasing $\tilde{\mu}$ makes the frequency increasingly dominated by the mass term, so the radial carrier represents a smaller fraction of the total phase scale of the packet. Equivalently, the characteristic radial transport associated with the carrier becomes less efficient at low $k_0$. In the Madelung description, the accretion rate is determined by the radial Noether flux, whose phase
content depends on the radial gradient $k_\ell$ together with the temporal
projection encoded in $\omega_\ell$. When the mass term dominates and the
initial radial carrier is small, the flux through the horizon over the finite
evolution time does not exhaust the exterior field. The result is the
low-$k_0$ partial-accretion floor seen in the maps in this regime $R_s > \lambdabar_C$.

As $k_0$ increases, the radial phase gradient imposed by the carrier becomes
larger, the transport of Noether charge toward the horizon becomes more
efficient, and the sectors that were partially accreted at low $k_0$ move
toward $\eta_{\ell,\mathrm{acc}} \simeq 1$. In this regime, increasing $k_0$
therefore does more than overcome the angular barrier, it also restores
efficient radial transport for packets whose low-$k_0$ dynamics is strongly
affected by the mass scale.

\begin{figure}[b]
  \centering
  \includegraphics[width=\columnwidth]{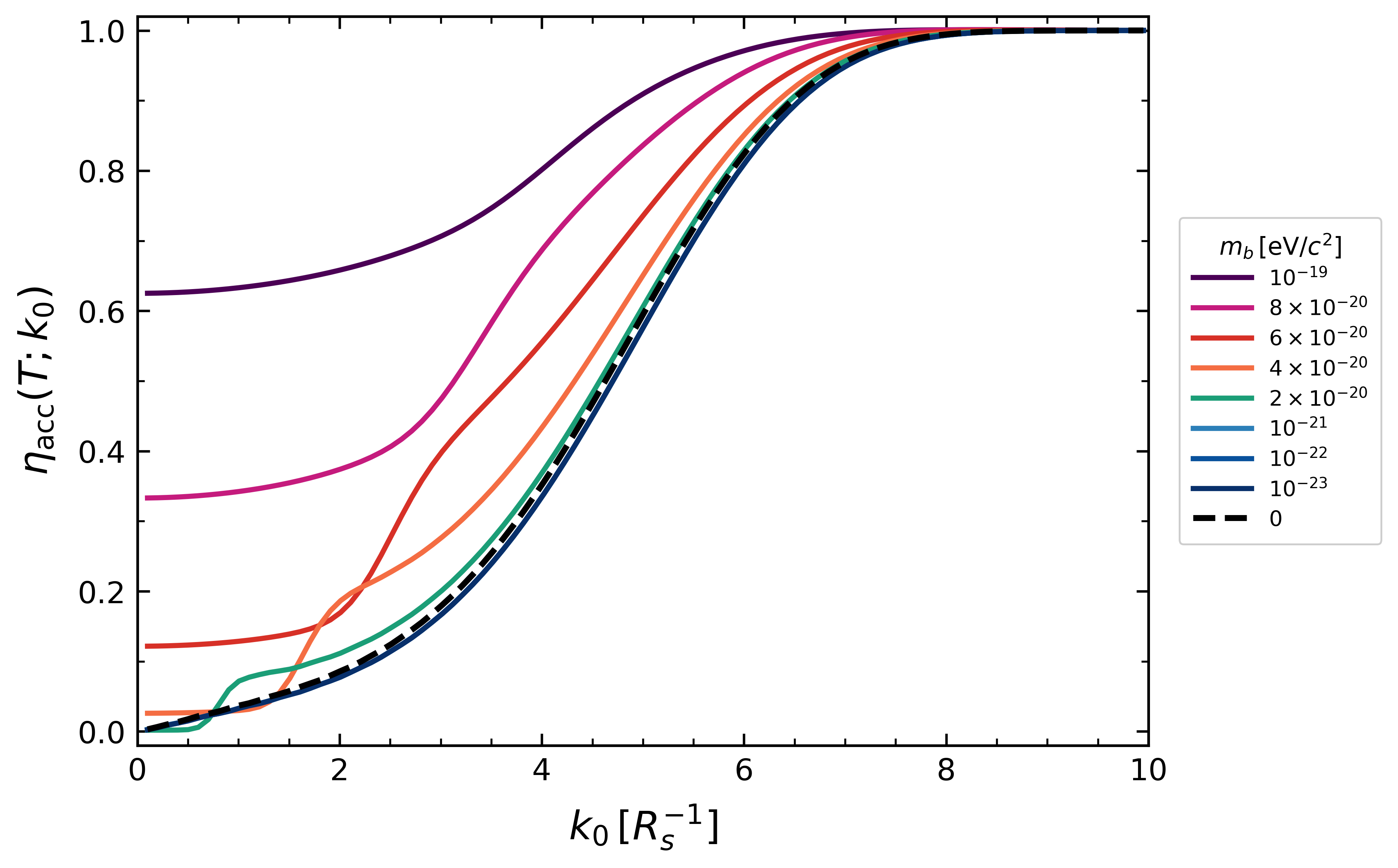}
  \caption{
  Summed accreted fraction over the representative $m=0$ sectors,
  $\eta_{\mathrm{acc}}(T;k_0)$, as a function of the wavenumber
  $k_0$ for the boson masses considered. Each curve corresponds to a different
  mass and represents the modal Noether-charge loss summed over the evolved
  representative sectors.  }
  \label{fig:eta_total}
\end{figure}

{\it Integrated representative accretion efficiency}. The modal maps can be condensed into a summed accreted fraction over the representative $m=0$ sectors by adding their Noether-charge losses. This quantity is not a degeneracy-weighted sum over all azimuthal modes, it is the integrated response of the same representative sectors used in the maps and axisymmetric reconstructions. For each $k_0$, we define

\begin{equation}
  \eta_{\mathrm{acc}}(T;k_0)  =  \frac{  \sum_{\ell=0}^{\ell_{\max}}  \int_0^T  \dot N_{\ell,\mathrm{acc}}(t;k_0)\,dt  }{  \sum_{\ell=0}^{\ell_{\max}} N_\ell(0;k_0)  } ,  \nonumber 
\end{equation}

\noindent which is shown in Fig. \ref{fig:eta_total} 
for all boson mass values considered.
Notice that $\eta_{\mathrm{acc}}$ increases with $k_0$ and approaches
unity once the radial carrier scale is large enough for most relevant
multipoles to be efficiently accreted. The shape of each curve reflects the
distribution of modal transitions in the corresponding two-dimensional maps.
When the transition is spread over a broad range of values of $k_0$, the integrated curve rises gradually. When many multipoles move from partial accretion to efficient
accretion within a narrow interval of $k_0$, the integrated curve rises more
steply.

The low-$k_0$ offset visible for the large masses is the global counterpart
of the partial-accretion floor in Fig. \ref{fig:Rs_ge_lambda}. It represents
the cumulative contribution of many modes with moderate accretion efficiency.
Thus, the summed curve should be read together with the maps, that is, a nonzero value of $\eta_{\mathrm{acc}}$ at small $k_0$ indicates partial charge
absorption across several representative sectors, while the approach to
$\eta_{\mathrm{acc}} \simeq 1$ indicates that almost all evolved sectors
carrying significant charge have entered the efficient accretion regime.

The accretion maps and the integrated curves show two qualitatively different
responses. For $R_s\leq\lambdabar_C$, the dynamics is nearly insensitive to the
boson mass over the range considered, and the classical barrier threshold
organizes the transition between reflected and efficiently accreted sectors.
For $R_s>\lambdabar_C$, the mass scale changes the character of the response,
low-$k_0$ packets enter a partial-accretion regime, the transition becomes weakly dependient in $\ell$, and efficient accretion is recovered as the radial carrier
scale increases. This change across $R_s=\lambdabar_C$ is the main result encoded
in the parameter-space maps.

\section{Discussion and conclusions}
\label{sec:conclusions}

We have studied the finite-time accretion of complex massive KG wave packets by
a Schwarzschild black hole in the test-field regime. The global $U(1)$ symmetry
of the complex field provides a conserved Noether current, and the associated
charge balance that we use to measure how much scalar matter crosses the
horizon and how much is scattered back into the exterior. In our analysis we use 
the initial conditions of nearly monochromatic Gaussian packets, radially localized with a radial wavenumber $k_0$, providing a simple way to prescribe localized scalar content with a well defined radial scale.

We exploited the spherical symmetry of the Schwarzschild background to reduce the
KG dynamics to independent radial evolution systems for the different multipolar
components of the initial scalar field. Because the reduced equations are independent of $m$, the simulations use the arbitrarily representative sector, $m=0$,
for each $\ell$. This allowed us to use a very efficient numerical method that solves 1+1 evolution problems that use small computer resources than axisymmetric evolutions, and in turn helped us to produce evolutions with high resolution within convergence regimes, explore a wide parameter space $(k_0,\ell)$ and produce detailed accretion maps.

We found that for fixed $k_0$, different multipoles separate into efficiently accreted, partially accreted, and mostly scattered sectors because they experience different angular barriers. The more efficiently accreted modes are those with low-$\ell$, while the less absorbed modes have higher-$\ell$. As a result, the late-time exterior field develops a more pronounced angular nodal structure. This shows that selective modal accretion modifies not only the radial profiles of the modes, but also the morphology of the scalar configuration that remains outside the black hole.

The accretion-efficiency maps in the $(k_0,\ell)$ plane are response diagrams of the black hole. Each point in these maps encodes the final fraction of initial 
Noether charge absorbed by the horizon. The classical transmission threshold
obtained from the maximum of the massive scalar effective potential follows the
main separation between exterior retention and efficient accretion. 
This agreement shows that, for $R_s \leq \lambdabar_C$,
the finite-time loss of Noether charge retains the organizing
structure expected from the standard barrier-transmission picture.

We produced these maps for various values of the scalar field mass and find that the response separates naturally according to the hierarchy between the Schwarzschild radius $R_s$ and the reduced Compton wavelength $\lambdabar_C$. For $R_s\leq\lambdabar_C$, the transition is broad, curved, and strongly organized by the angular barrier. In this regime, the boson mass produces only mild quantitative changes over the range considered, and the value of $k_0$ required for efficient accretion increases systematically with $\ell$. For $R_s>\lambdabar_C$, the mass scale changes the character of the response and  the transition becomes more collective across multipoles, and interestingly a partial-accretion
floor appears at low $k_0$.

The low-$k_0$ partial-accretion floor has a direct interpretation in terms of
the phase content of the packet. When the mass term dominates the asymptotic
dispersion relation, the imposed radial carrier represents a smaller fraction
of the total phase scale, and the radial Noether flux does not exhaust the
exterior field over the finite evolution time. Increasing $k_0$ restores
efficient radial transport and drives the corresponding sectors toward
$\eta_{\ell,\mathrm{acc}}\simeq1$. In this sense, the accretion maps provide a
charge-based counterpart of the usual spectral picture, that is, the same effective
potential landscape that supports long-lived massive-scalar configurations also
determines which localized multipolar components remain outside the black hole
after a finite interaction time.

The integrated representative accretion efficiency condenses the modal maps into a single measure of scalar absorption over the evolved $m=0$ sectors. Its
dependence on $k_0$ reflects the distribution of modal transitions in the
two-dimensional response diagrams. Gradual increase in the integrated curve corresponds to broad modal transitions, while steeper rises appear when many multipoles move from partial accretion to efficient accretion over a narrower range of $k_0$. The nonzero low-$k_0$ offset for the largest masses is the global manifestation of the partial-accretion floor seen in the modal maps.

Overall, the results provide a finite-time, charge-based classification of
massive scalar-field accretion by a Schwarzschild black hole. The method
identifies which representative multipolar sectors of a localized wave packet
are effectively absorbed, which sectors remain outside, and how this division
is controlled by $k_0$, $\ell$, and the scalar-field mass.

We expect that the multimode nature of the wave-packets used in our analysis helps to quantify the multimode accretion of Fuzzy Dark Matter, since it has a granular and dynamical structures, not only in galactic halos, but also at the core as shown in condensation processis around black holes \cite{PalomaresChavez:2025BHCondensation}. Moreover, in this paper we have covered a wide range of scalar field mass that can be identified with its non-relativistic countyerpart.

Our approach considering the black hole as a filter of certain modes, leads to a rather curious conclusions, namely that the non-accreted remnant scalar field is characterized by high values of $\ell$, which indicates that perhaps scalar wigs \cite{Barranco2012ScalarWigs} should have a tendency to be dominated by high-$\ell$ terms.

\section*{Acknowledgments}
This research is supported by  
SECIHTI Grant No. CFB-2025-I-759, 
Laboratorio Nacional de C\'omputo de Alto Desempe\~no Grant No. 2026-8,
CIC-UMSNH Grant No. 4.9.

\section*{Data AVAILABILITY}
The data produced for our analysis are openly available at \cite{rosales_infante_2026_20130371}.

\bibliography{bibliography}

\appendix

\section{Numerical validation} \label{app:numerical_validation}

This appendix presents a convergence test of the evolution scheme. For this we use the wave-packet in Eqs.~\eqref{eq:initial_phi}--\eqref{eq:initial_pi},
including the same physically motivated choices for $r_0$, $\sigma$, and final
time. The central wavenumber is fixed to a representative value $k_0 = 0.5\,R_s^{-1}$.

We use radial discretization using $N_r = 12000,18000,27000$ cells, or equivalently three resolutions $\Delta r = (600 - 0.5)R_s / Nr$. We then perform a Richardson self-convergence analysis using the Noether charge $N(t)$ \eqref{eq:Noether_charge}. This procedure follows standard practices for estimating discretization errors from multiple resolutions \cite{Celik2008,Guzman2023}.

Since the factor between consecutive resolutions is $1.5$, the self-convergence exponent can be computed as

\begin{equation}
  p(t) = \log_{1.5}\!\left(
    \frac{N_{12000}(t)-N_{18000}(t)}{N_{18000}(t)-N_{27000}(t)}
  \right).\nonumber
\end{equation}

\begin{figure}[t]
  \centering
  \includegraphics[width=0.49\textwidth]{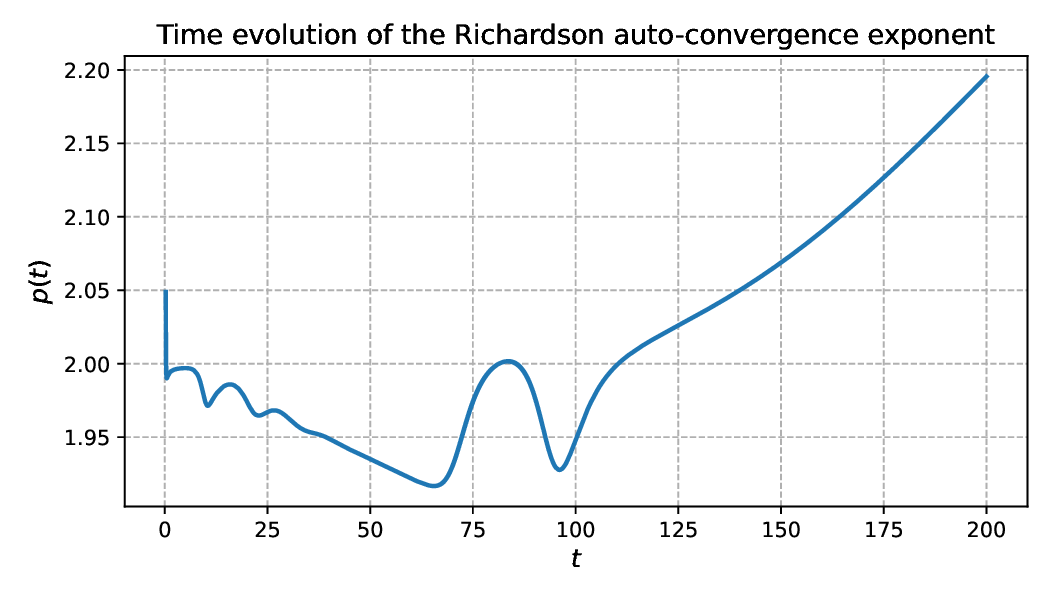}
  \caption{Time evolution of the Richardson self-convergence exponent $p(t)$ computed
    from the Noether charge using three resolutions $N_r= 12000,18000,27000$.
    Values close to $p\simeq 2$ indicate second-order self-convergence during the evolution.}
  \label{fig:auto_convergence}
\end{figure}

\noindent Figure \ref{fig:auto_convergence} shows the resulting time-dependent exponent
$p(t)$. After a short initial transient, $p(t)$ remains close to $p\simeq 2$
over most of the evolution, with moderate time variability associated with the
wave-packet dynamics. This behavior is consistent with the nominal second-order
accuracy of the radial finite difference discretization, that dominates over the third order of the evolution integrator RK3, and validates the consistency of our implementation.

As an additional validation, we test the global Noether-charge
budget implied by the flux-balance form of the continuity equation
\eqref{eq:continuity}. Since particle number can leave a finite radial domain
through its boundaries, the appropriate diagnostic is not the constancy of
$N(t)$ itself, but the consistency of its evolution with the net flux of the
Noether current. For this, we define the conservation defect

\begin{equation}
  D(t)\equiv N(t)-N(0)-\int_0^t F_{\mathrm{net}}(t')\,dt',
  \nonumber \label{eq:conservation_defect}
\end{equation}

\noindent where $F_{\mathrm{net}}(t)$ denotes the net radial flux through the domain
boundaries, computed from the discrete evaluation of $\alpha\sqrt{\gamma}\,J^r$.
For a globally consistent discretization of both $N(t)$ and the boundary fluxes,
one expects $D(t)\to 0$ as $\Delta r \to 0$.

\begin{figure}[t]
  \centering
  \includegraphics[width=\columnwidth]{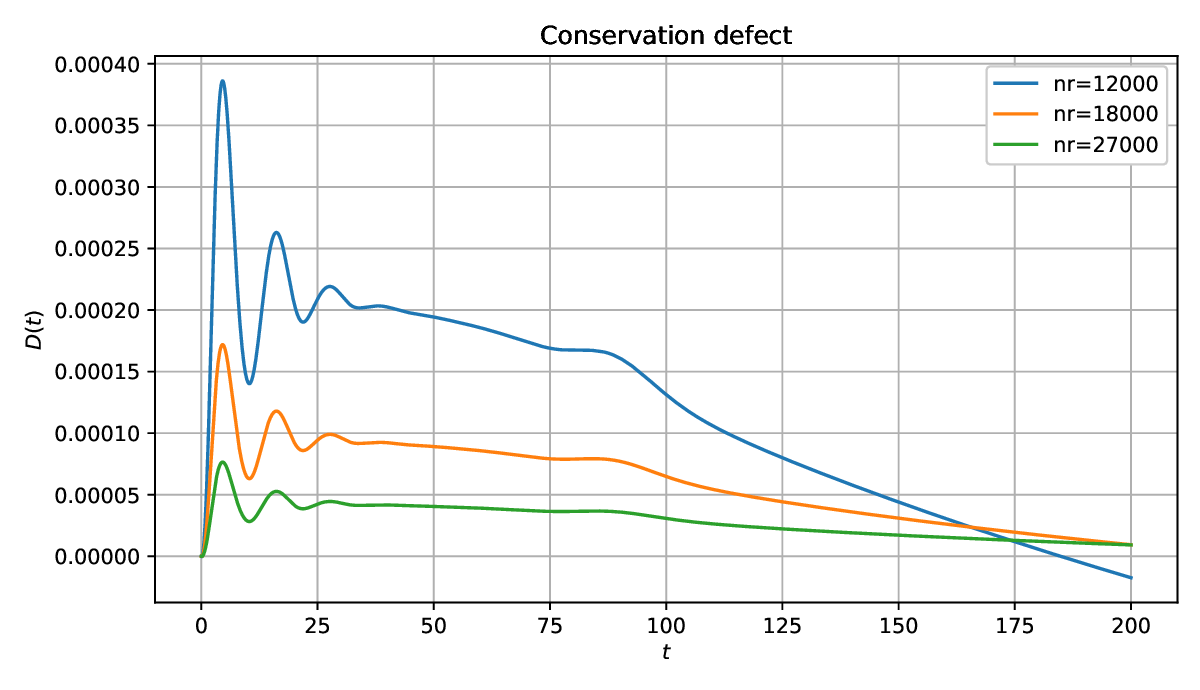}
  \includegraphics[width=\columnwidth]{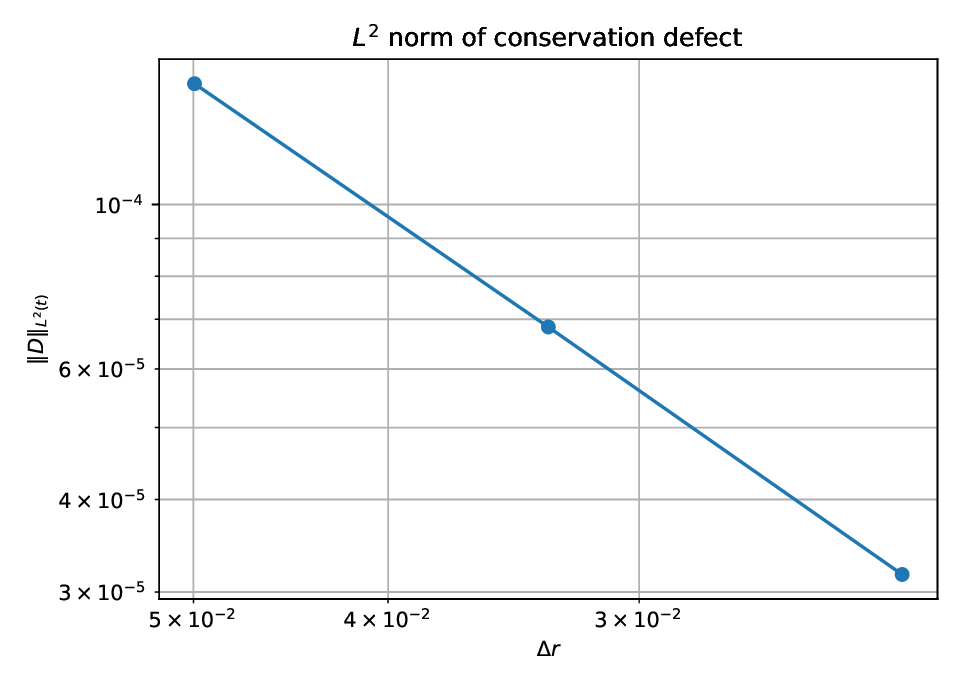}
  \caption{Convergence of the global Noether-charge budget.
    (Top) Time evolution of the conservation defect $D(t)$ defined in
    Eq.~\eqref{eq:conservation_defect} for three radial resolutions
    with $N_r=12000,18000,27000$.
    (Bottom). The value of $\|D\|_{L^2(t)}$ as a function of resolution
  $\Delta r$, showing consistent decay of the error with resolution.}
  \label{fig:noether_defect}
\end{figure}

The top panel of Fig. \ref{fig:noether_defect} shows the time evolution of $D(t)$
for the three resolutions. Initially there is an early-time transient, but its
magnitude decreases with increasing resolution, indicating that
the global Noether budget is recovered under grid refinement. To quantify this
trend, we compute the $L^2$ norm

\begin{equation}
  \|D\|_{L^2(t)}=\left(\int_0^{T} |D(t)|^2\,dt\right)^{1/2},
\end{equation}

\noindent and estimate its dependence on the radial grid $\Delta r$. As shown in the bottom panel of Fig. \ref{fig:noether_defect}, $\|D\|_{L^2(t)}$ decreases
as $\Delta r$ is reduced, providing direct evidence that
violations of the global conservation law converge to zero in the continuum
limit, as expected for consistency
\cite{Oberkampf2002,Guzman2023}.

{\it Inner boundary location test.} As a further check on the influence of the inner boundary location, we repeat the reference simulation with five values of $r_{\min}/R_s = 0.5, 0.6, 0.7, 0.8, 0.9$, all placed inside the event horizon.

\begin{figure}[t]
    \centering
    \includegraphics[width=\columnwidth]{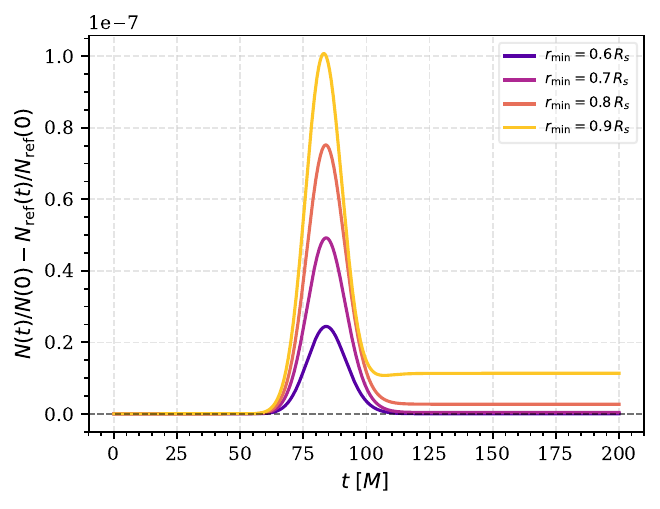}
    \caption{Difference in the normalized Noether charge $\Delta N(t)$
    between simulations with inner boundary at
    $r_{\min}/R_s = 0.6, 0.7, 0.8, 0.9$ and the case of production runs with
    $r_{\min}/R_s = 0.5$, for $k_0 = 0.5\,R_s^{-1}$.
    The deviations are at the level of $|\Delta N| \lesssim \mathcal{O}(10^{-7})$, in the lapse when the pulse interacts with the black hole. This test 
    shows that the inner boundary location does not introduce measurable contamination on the evolution outside the black hole where diagnostics is carried out.}
    \label{fig:rmin_test}
\end{figure}

Figure \ref{fig:rmin_test} shows the difference in the normalized Noether charge,

\begin{equation}
    \Delta N(t) \equiv \frac{N(t)}{N(0)}\bigg|_{r_{\min}}
    - \frac{N(t)}{N(0)}\bigg|_{r_{\min}=0.5\,R_s},
\end{equation}

\noindent with respect to the value used in production runs at $r_{\min} = 0.5\,R_s$.
The deviations remain at the level of $|\Delta N| \lesssim \mathcal{O}(10^{-7})$ throughout the entire evolution, confirming that the inner boundary has no
measurable effect on the exterior dynamics and that the outgoing characteristic
condition at $r_{\min}$ prevents measurable spurious reflection from inside the black hole.

\end{document}